\documentclass[twocolumn]{aastex63}
\usepackage{amssymb}	
\usepackage{amsmath}	
\usepackage{array}
\usepackage{bm}
\usepackage{CJK}
\usepackage{graphicx}	
\usepackage{hyperref}
\usepackage{xifthen}
\usepackage{xspace}
\usepackage{booktabs}
\usepackage{here}

\begin{document}
\begin{CJK*}{UTF8}{gbsn}
\title[Spiral Pattern Speed in IM Lup]{No Significant Rotation of the Spiral Structure in the IM Lup Protoplanetary Disk}

\author{Yuto Hiraide}
\affiliation{College of Science, Ibaraki University, 2-1-1 Bunkyo, Mito, 310-8512, Ibaraki, Japan }

\author[0000-0001-8877-4497]{Masataka Aizawa}
\affiliation{College of Science, Ibaraki University, 2-1-1 Bunkyo, Mito, 310-8512, Ibaraki, Japan }

\author[0000-0002-3001-0897]{Munetake Momose}
\affiliation{College of Science, Ibaraki University, 2-1-1 Bunkyo, Mito, 310-8512, Ibaraki, Japan }
\affiliation{NAOJ Chile, National Astronomical Observatory of Japan, Los Abedules 3085 Oficina 701, Vitacura 7630414, Santiago, Chile}

\begin{abstract}
Spiral structures are often observed in protoplanetary disks, but their origins remain uncertain.
\citet{Yoshida2025} reported a pattern speed consistent with local Keplerian rotation for the $m=2$ spiral in the dusty disk around IM~Lup using multi-epoch ALMA observations, supporting a gravitational instability (GI)-driven origin.
However, recent studies revise the disk geometry of IM~Lup, motivating a reassessment of its spiral pattern speed.
In this study, we reanalyze the disk geometry of the same multi-epoch data using visibility-based continuum  emission modeling.
The inferred disk geometries differ from those adopted in previous studies by a few degrees but are consistent across epochs and with recent independent estimates based on gas emission analyses. The revised geometries yield weaker residuals and more physically plausible radial spiral phase profiles.
Unlike previous studies, our remeasured spiral pattern speeds are consistent with zero rotation within the uncertainties.
We conclude that the previously reported Keplerian-like motion can be explained by a mismatch in the adopted disk geometry, which affects the spiral phases measured in Band~6 (2017, 2019) and Band~7 (2024) differently.
In addition, we identify a tentative large-scale one-armed spiral-like structure extending beyond the known $m=2$ spiral, which may provide clues to additional perturbations, such as an embedded planet or late-stage infall. Future high-resolution Band~6 observations with a longer time baseline will be essential for more accurately determining the spiral pattern speed.

\end{abstract}

\section{Introduction}

High-angular-resolution observations with the Atacama Large Millimeter/submillimeter Array (ALMA) have revealed that protoplanetary disks exhibit a wide variety of substructures. The Disk Substructures at High Angular Resolution Project (DSHARP) systematically observed 20 Class~II protoplanetary disks in millimeter dust continuum emission and demonstrated that substructures such as rings, gaps, and spirals are common in protoplanetary disk \citep[e.g.,][]{Andrews2018}. In particular, spiral structures are of great interest because they may trace disk dynamical processes related to planet formation and angular momentum transport \citep[e.g.,][]{Dong2018,Longarini2026}.

Although spiral structures in protoplanetary disks have been observed in near-infrared scattered-light \citep[e.g.,][]{Fukagawa2004,Muto2012,Grady2013}, millimeter dust continuum emission \citep[e.g.,][]{Huang2018b,Lee2020,Huang2025} and molecular emission \citep[e.g.,][]{Huang2021,Speedie2025}, the origin of spiral structures remains uncertain.
Planet-disk interactions \citep[e.g.,][]{Fung2015,Bae_Zhu2018} and gravitational instability (GI) \citep[e.g.,][]{Kratter2016,Longarini2026} are the two leading mechanisms discussed for producing spirals. In addition, other scenarios such as shadow-induced temperature perturbations \citep[e.g.,][]{Montesinos2016,Su2024}, stellar flybys \citep[e.g.,][]{Cuello2019,Su2026}, and spirals driven by interactions between misaligned inner and outer disk regions \citep{Rowther2025} are proposed.

Several observational diagnostics are explored to constrain the origin of disk spirals, including disk-mass and gravitational-stability estimates \citep[e.g.,][]{Kratter2016,Longarini2026}, spiral morphology \citep[e.g.,][]{Fung2015,Forgan2018,Bae2018b}, and molecular line kinematic signatures of gravitational instability or embedded planets \citep[e.g.,][]{Perez2015,Hall2020}.
Recently, spiral pattern speed measurements are emerged as a more direct observational diagnostic of spiral origins \citep[e.g.,][]{Ren2020,Xie2024,Ren2024}.
Planet-driven spirals are expected to rotate nearly rigidly at the orbital angular velocity of the perturbing planet \citep[e.g.,][]{Ogilvie2002,Bae_Zhu2018}, whereas GI-driven spirals are predicted to approximately follow the local Keplerian rotation of the disk \citep[e.g.,][]{Cossins2009,Bethune2021}.

The target of this study is a Class~II protoplanetary disk around the young T~Tauri star IM~Lup, located at a distance of 155.8~pc \citep{Gaia2023}, which exhibits a prominent $m=2$ spiral structure \citep{Huang2018b}.
Using multi-epoch ALMA observations, \citet{Yoshida2025} captured the winding motion of the IM~Lup dust spiral and found that its pattern speed is consistent with local Keplerian rotation, supporting a GI-driven origin and suggesting that the observed spirals may be transient features.
This interpretation is also consistent with the relatively high disk mass of IM~Lup, whose central star has a mass of $M_{\star}\sim1.1\,M_{\odot}$ and whose disk-to-star mass ratio is of order $\sim0.1$ \citep{Teague2021,Lodato2023,Martire2024}, for which disk self-gravity can become dynamically important and excite spiral density wave \citep[e.g.,][]{Kratter2016,Tomida2017,Veronesi2021,Pezzotta2026}.
Furthermore, \citet{Ueda2024} proposes that a gravitationally self-regulated disk model can provide a unified explanation for the multi-wavelength properties of IM~Lup.

On the other hand,
a planet-driven origin is proposed for the spirals.
\citet{Pinte2020} reports a CO velocity kink, which is interpreted as a possible kinematic signature of an embedded planet \citep{Perez2015}.
The inferred planet location is suggested to be associated with the gap at $\sim117$~au.
\citet{Verrios2022} further proposes that a single giant planet at $\sim117$~au can provide a unified explanation for the CO velocity kink \citep{Pinte2020}, the millimeter continuum spirals \citep{Huang2018b}, and the spiral-like substructures seen in near-infrared scattered-light \citep{Avenhaus2018}.

Measuring spiral morphology in millimeter dust continuum emission requires accurate subtraction of the axisymmetric disk component using well-constrained disk geometry \citep{Aizawa2024}.
A suitable tool for this purpose is the open-source package \texttt{protomidpy} \citep{Aizawa2024}, which provides a visibility-based method that simultaneously estimates the one-dimensional radial brightness profile and disk geometry, including the disk center, inclination, and PA.
Indeed, for IM~Lup, \citet{Aizawa2026a} find that the disk geometry inferred with \texttt{protomidpy} differs by a few degrees from that adopted in \citet{Yoshida2025} from \citet{Huang2018a}, and that this difference affects the extracted spiral morphology. 
We therefore apply \texttt{protomidpy} to the four ALMA datasets used by \citet{Yoshida2025}, construct axisymmetric disk models and residual images for each epoch, and remeasure the spiral pattern speed using the updated geometries.

This paper is structured as follows. Section~\ref{sec2} describes the data and the methods used to generate residual images and measure the spiral pattern speed. Section~\ref{sec3} shows the results of \texttt{protomidpy} analysis and residual images.  Section~\ref{sec4} presents the spiral morphology and the measured pattern speed profiles by Fourier analysis. Section~\ref{sec5} discusses the implications of our results, and Section~\ref{sec6} summarizes our conclusions.

\section{Observations and Method}
\label{sec2}

\subsection{data}
\label{subsec:data}
We analyze the same calibrated measurement sets of the IM~Lup protoplanetary disk as those used by \citet{Yoshida2025}.
A summary of the observations is given in Table~\ref{tab:obs_summary}.
The data comprise four ALMA observations obtained over a span of roughly seven years, which we refer to as the 2017, 2019a, 2019b, and 2024 epochs.
The 2017 epoch is based on the DSHARP Band~6 data with additional short-baseline archival observations \citep{Andrews2018}.
The 2019a and 2019b epochs use the Band~6 MAPS project \citep{Oberg2021} data obtained with two different frequency settings.
Following \citet{Yoshida2025}, we treat the two MAPS frequency settings as separate epochs.
The 2024 epoch is based on the Band~7 data presented by \citet{Yoshida2025}.

To apply \texttt{protomidpy}, we preprocess the calibrated continuum visibilities following \citet{Aizawa2024}.
Each spectral window is averaged into a single channel, and a 30~s time averaging is applied to reduce the data volume.
Because the recorded weights in the measurement sets can be systematically overestimated \citep{Hashimoto2021,Aizawa2024}, we manually rescale the visibility weights following \citet{Aizawa2026a} by dividing them by correction factors of order a few. The adopted factors are listed in Table~\ref{tab:params} as the Weight factor.

\begin{table*}[t]
  \centering
  \caption{Summary of IM~Lup ALMA observations.}
  \label{tab:obs_summary}
  
  \begin{tabular*}{\textwidth}{@{\extracolsep{\fill}}lcccccc}
    \toprule
    Epoch & Date [MJD] & Frequency (GHz) & Band & uv range (k$\lambda$) & Project ID & Reference \\
    \midrule
    2017  & 2017 Oct 10 [58036] & 238.984 & 6 & 12 -- 11976 & 2016.1.00484.L & (1) \\
    2019a & 2019 Aug 13 [58708] & 226.053 & 6 & 11 -- 2743  & 2018.1.01255.L & (2) \\
    2019b & 2019 Aug 20 [58715] & 257.029 & 6 & 13 -- 2912  & 2018.1.01255.L & (2) \\
    2024  & 2024 Mar 22 [60391] & 330.172 & 7 & 16 -- 2772  & 2023.1.00525.S & (3) \\
    \bottomrule
  \end{tabular*}
  
   \vspace{3pt}
  \begin{flushleft}
    {\small
    \textit{Note.}
    This table is adapted from Table~1 of \citet{Yoshida2025}.
    The $uv$ range is the baseline length normalized by the observing wavelength, $\lambda$.
    The 2017 epoch includes the DSHARP data and additional short-baseline observations obtained in 2014--2015.
    References: (1) \citet{Andrews2018}; (2) \citet{Oberg2021}; (3) \citet{Yoshida2025}.}
  \end{flushleft}
\end{table*}

\subsection{Modeling of axisymmetric geometrically thin disk}
\label{subsec:Modeling}
For each epoch, we model the millimeter continuum emission as a geometrically thin, axisymmetric disk using \texttt{protomidpy} \citep{Aizawa2024}.
In \texttt{protomidpy}, the disk is described by a one-dimensional radial intensity profile and a set of geometric parameters.
The radial intensity profile is discretized as
\begin{equation}
    \mathbf{a} = \{ I(r_k) \}_{k=1}^{N},
\end{equation}
where $I(r_k)$ is the intensity at the radial collocation points, defined as $r_k = j_{0,k}R_{\rm out}/j_{0,N+1}$, with  $j_{0,k}$ representing the k-th root of the Bessel function $J_0$, and $R_{\rm out}$ the outer boundary of the disk.
The disk geometry is parameterized as
\begin{equation}
    \mathbf{g} =
    (\Delta x_{\rm cen}, \Delta y_{\rm cen}, \cos i, {\rm PA}),
\end{equation}
where $(\Delta x_{\rm cen}, \Delta y_{\rm cen})$ are the offsets of the disk center in the sky plane, $i$ is the inclination, and PA is the position angle of the disk major axis measured counterclockwise from north.
The inclination is defined as the angle between the line of sight and the disk rotation axis, so that $\cos i$ corresponds to the apparent axis ratio of a geometrically thin disk.

To regularize the radial intensity profile and prevent overfitting, \texttt{protomidpy} models the radial correlations using a Gaussian process with a radial basis function kernel,
\begin{equation}
    K(r,r') =
    \alpha \exp\left[-\frac{(r-r')^2}{2\gamma^2}\right],
\end{equation}
where $\alpha$ controls the weight of the prior and $\gamma$ controls the radial correlation length.
The hyperparameters are therefore given by
\begin{equation}
    \boldsymbol{\theta} = (\alpha,\gamma).
\end{equation}
A key advantage of \texttt{protomidpy} is that it simultaneously estimates the radial intensity profile, disk geometry, and hyperparameters directly from the observed visibilities.
This allows the disk geometry to be inferred without relying on CLEAN images, which are secondary products of interferometric data. 
It allows extraction of subtle features of the disk, such as a spiral with an amplitude much smaller than the angle-averaged intensities, or the brightness temperature in the innermost regions of the disk \citep{Aizawa2026a}.
The related visibility-based code \texttt{frank} \citep{Jennings2020} reconstructs axisymmetric radial brightness profiles but requires the disk geometry and hyperparameters to be fixed in advance.
These fixed choices may introduce manual biases, and if they are not optimal, the resulting residual images may contain artificial non-axisymmetric structures \citep[e.g.,][]{Andrews2021,Jennings2022a,Aizawa2024}.

In constructing the radial intensity profile, we use $N=300$ collocation points and set the outer boundary to $R_{\rm out}=3^{\prime\prime}$.
\texttt{protomidpy} infers the posterior distributions of the geometric parameters and hyperparameters using the affine invariant Markov Chain Monte Carlo (MCMC) code \texttt{emcee}.
We adopt uniform priors for $\cos i$, PA, the disk center, and $\gamma$, and a log-uniform prior for $\alpha$.
We use 32 walkers and run the sampler for 10000 steps, discarding the first 3000 step as burn-in. We assess convergence using the integrated autocorrelation time, following \citet{Aizawa2024}, and confirmed that the chains are sufficiently long for all parameters.

The MAP estimates of the geometric parameters and hyperparameters, together with the corresponding best-fit radial intensity profile, are adopted as the representative axisymmetric model for each epoch.
We then obtain the residual visibilities by subtracting the model visibilities of the axisymmetric disk from the observed visibilities.
\subsection{Making residual images}
\label{subsec:residual_imaging}

We create residual images following \citet{Aizawa2024,Aizawa2026a} from the residual visibilities.
Using the disk geometry inferred with \texttt{protomidpy}, we generate residual images both in the observed frame and in the deprojected flame, where the sky-plane coordinates are transformed to the disk plane using the inferred inclination and position angle (see \cite{Aizawa2024} for details).
The deprojected residual images are used to analyze the spiral morphology and measuring pattern speed of the spiral.

We produce the residual images using the CASA task \texttt{tclean}.
The images are made with a pixel scale of 6~mas and an image size of $1000 \times 1000$ pixels.
We adopt the multiscale CLEAN algorithm with scale sizes of $[0, 30, 120, 360, 720, 1440]~{\rm mas}$ a CLEAN threshold of $3.5\sigma$. 
These imaging parameters follow those adopted in \citet{Aizawa2024,Aizawa2026a}.
We also construct even- and odd-symmetric residual images by separately imaging the real and imaginary components of the residual visibilities.
The real-only images isolate even-symmetric residuals such as the $m=2$ mode, whereas the imaginary-only images reveal odd-symmetric structures such as the $m=1$ mode.
For the pattern speed analysis, we prepare deprojected residual images using imaging parameters chosen to be as close as possible to those of \citet{Yoshida2025}.
The 2017 data are imaged with a \texttt{uv}-taper of 0.08~arcsec and super-uniform weighting with \texttt{npixels = 10}.
The 2019a, 2019b, and 2024 data are imaged with Briggs robust parameters of $-0.2$, 0.0, and 0.0, respectively.
All deprojected residual images are then convolved to a common circular beam with an FWHM of 150~mas.

For comparison, we repeat the residual imaging and pattern speed analysis using the inclination and PA adopted in \citet{Yoshida2025} from \citet{Huang2018a}, while fixing the disk center to the \texttt{protomidpy} value in both cases, since the disk center offset is also accounted for in \citet{Yoshida2025}.

\subsection{Fourier analysis of spiral arm}
\label{subsec:fourier_analysis}

We analyze the spiral morphology and the pattern speed from the deprojected residual images.
To quantify the spiral morphology, we follow the Fourier analysis used by \citet{Aizawa2026a}.
For each epoch, we evaluate the deprojected image on a polar grid and decompose the azimuthal brightness profile at each radius into Fourier modes as
\begin{equation}
    I(r,\phi)
    =
    \sum_{m=0}^{\infty}
    I_m(r)
    \cos\left[
        m(
            \phi - \phi_m(r)
        )
    \right],
    \label{eq:fourier_decomp}
\end{equation}
where $m$ is the azimuthal mode number, $I_m(r)$ is the amplitude of the $m$-th mode, and $\phi_m(r)$ is its phase.
In practice, we sample the azimuthal profile at each radius with $N_\phi$ uniformly spaced points,
$\phi_k = 2\pi k/N_\phi$ $(k = 0, 1, \dots, N_\phi-1)$,
and compute the complex Fourier coefficient
\begin{equation}
    C_m(r)
    =
    \sum_{k=0}^{N_\phi-1}
    I(r,\phi_k)
    \exp(-im\phi_k).
    \label{eq:complex_fourier_coeff}
\end{equation}
For the non-axisymmetric modes ($m \geq 1$), the amplitude and phase
of the $m$-th mode are given by
\begin{align}
    I_m(r)
    &=
    \frac{2}{N_\phi}
    \left| C_m(r) \right|,
    \label{eq:fourier_amp}
    \\
    \phi_m(r)
    &=
    -\frac{1}{m}
    \arg \left[ C_m(r) \right].
    \label{eq:fourier_phase}
\end{align}
Since the IM~Lup disk is dominated by a roughly symmetric two-armed spiral, we focus on the $m=2$ component.

At each radius, we measure the phase $\phi_2(r,t_i)$ for each epoch $t_i$ and trace its temporal evolution.
Assuming that the pattern speed remains constant at each radius, we estimate $\omega(r)$ from a linear fit to $\phi_2(r,t_i)$ as a function of time.
The uncertainty in $\omega(r)$ is estimated by propagating the phase uncertainties through a weighted linear regression; the details are described in Appendix~\ref{app:Uncertainty}. These uncertainties represent only statistic errors, and  systematic errors not accounted in the current analysis may further increase the uncertainties.
\section{Axisymmetric Model and Residual Images}
\label{sec3}

\subsection{Disk geometry and 1D Intensity profile}
\label{subsec:modeling_result}
\begin{table*}[htbp]
\centering
\small
\caption{Disk parameters and Weight factors.
}
\label{tab:params}
\begin{tabular}{lcccccccc}
\toprule
Epoch &
$\Delta x_{\rm cen}$ (mas) &
$\Delta y_{\rm cen}$ (mas) &
$\cos i$ &
$i$ (deg) &
PA (deg)&
$\gamma$ (arcsec) &
$\log_{10}\alpha$ &
Weight factor
 \\
\midrule
Huang+2018 &
 $-1.5^{+2}_{-2}$ &
 $1^{+2}_{-2}$ &
 $0.6756^{+0.0039}_{-0.0039}$ &
 $47.5^{+0.3}_{-0.3}$ &
 $144.5^{+0.5}_{-0.5}$ &
 $-$ &
 $-$ &
 $-$ \\
Paneque+2024(CN)&
 $-$ &
 $-$ &
 $0.6356$ &
 $50.53$ &
 $143.76$ &
 $-$ &
 $-$ &
 $-$ \\
2017 &
 $1.125^{+0.051}_{-0.051}$ &
 $-0.700^{+0.052}_{-0.049}$ &
 $0.6378^{+0.0005}_{-0.0006}$ &
 $50.37^{+0.04}_{-0.04}$ &
 $143.83^{+0.06}_{-0.06}$ &
 $0.077^{+0.003}_{-0.005}$ &
 $2.829^{+0.211}_{-0.336}$ &
 $2.8455$ \\
2019a &
 $-3.478^{+0.062}_{-0.062}$ &
 $-0.976^{+0.057}_{-0.056}$ &
 $0.6321^{+0.0006}_{-0.0006}$ &
 $50.80^{+0.04}_{-0.04}$ &
 $143.61^{+0.06}_{-0.06}$ &
 $0.145^{+0.006}_{-0.006}$ &
 $1.158^{+0.219}_{-0.210}$ &
 $2.1401$ \\
2019b &
 $-1.154^{+0.056}_{-0.056}$ &
 $-1.621^{+0.053}_{-0.053}$ &
 $0.6343^{+0.0004}_{-0.0004}$ &
 $50.63^{+0.03}_{-0.03}$ &
 $144.36^{+0.04}_{-0.04}$ &
 $0.147^{+0.005}_{-0.005}$ &
 $1.526^{+0.205}_{-0.199}$ &
 $2.1178$ \\
2024 &
 $6.554^{+0.058}_{-0.057}$ &
 $4.723^{+0.055}_{-0.056}$ &
 $0.6436^{+0.0004}_{-0.0004}$ &
 $49.94^{+0.03}_{-0.03}$ &
 $143.72^{+0.04}_{-0.04}$ &
 $0.121^{+0.005}_{-0.005}$ &
 $0.559^{+0.158}_{-0.149}$ &
 $3.5057$ \\
\bottomrule
\end{tabular}
\end{table*}

\begin{figure}[b]
    \centering
    \includegraphics[width=1.0\linewidth]{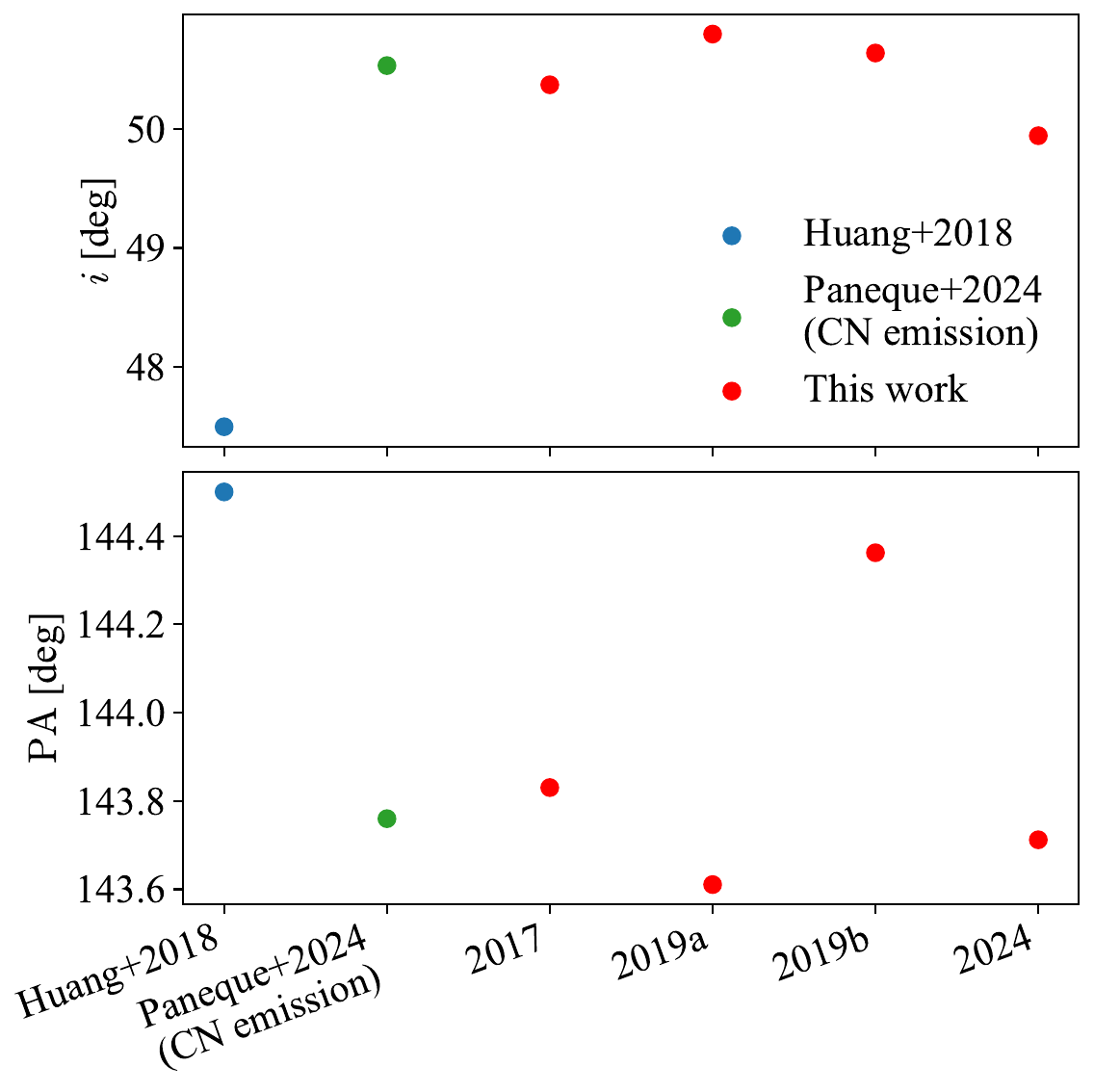}
\caption{Disk geometry of IM~Lup inferred in this work for each epoch, compared with reported by \citet{Huang2018a} and \citet{Paneque2024}. The top and bottom panels show the inclination and PA, respectively.
}
    \label{fig:inferred_geometry}
\end{figure}

\begin{figure*}[ht]
    \centering
    \includegraphics[width=0.95\linewidth]{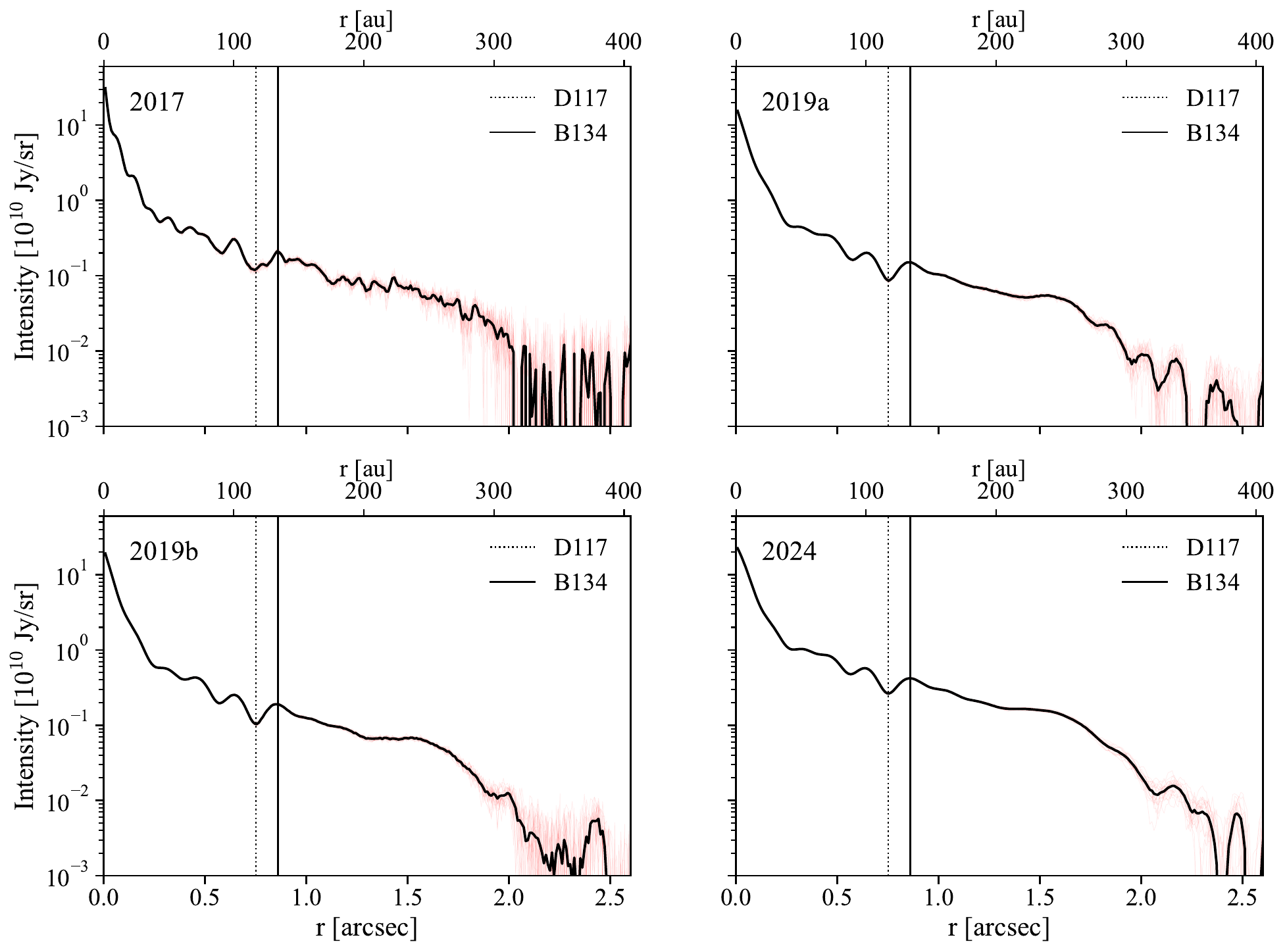}
    \caption{Radial intensity profiles of the axisymmetric disk model inferred with \texttt{protomidpy} for each epoch.
    The black curves represent the adopted radial intensity profiles, and the red curves show 20 samples from posterior, illustrating the uncertainty of the reconstructed profiles.
    The vertical dotted and solid lines mark the D117 gap and B134 ring reported by \citet{Huang2018a}, respectively.}
    \label{fig:1D_profiles}
\end{figure*}
Table~\ref{tab:params} summarizes the disk geometric parameters, hyperparameters, and weight factors estimated with \texttt{protomidpy} for each epoch.
Figure~\ref{fig:inferred_geometry} compares the disk inclination and PA inferred in this work at each epoch with those reported by \citet{Huang2018a} and \citet{Paneque2024}.
In the DSHARP analysis of the IM~Lup disk, these geometric parameters were estimated by fitting ellipses to annular substructures in the image plane, yielding $i = 47.5 \pm 0.3^\circ$ and ${\rm PA} = 144.5 \pm 0.5^\circ$ \citep{Huang2018a}. In this study, we obtain mean values of $i = 50.43^\circ$ and ${\rm PA} = 143.88^\circ$, with the values inferred independently for each epoch being broadly consistent with one another. Similar values of $i = 50.53^\circ$ and ${\rm PA} = 143.76^\circ$ were independently obtained by \citet{Paneque2024} through modeling of CN emission using \texttt{DISCMINER} \citep{Izquierdo2021}. The agreement despite the use of a different tracer and methodology provides independent support for the disk geometry inferred in this work. 
The inferred geometries differ systematically from the image-based geometry derived by \citet{Huang2018a}, by about $3^\circ$ in inclination and up to $1^\circ$ in PA.
The inferred geometry varies by less than 1 deg across epochs, with only minor effects on the residual structures (see Section \ref{residual_image})

Figure~\ref{fig:1D_profiles} shows the radial brightness profiles inferred for the four epochs.
Although the absolute intensities differ among epochs, the profiles show similar overall trends and recover the annular substructures reported by \citet{Huang2018a}, including the D117 gap and the B134 ring.
The 2017 profile is consistent with that reported by \citet{Aizawa2026a} and shows finer radial structures than the profiles of the other epochs.

\subsection{Residual images}
\label{residual_image}
\begin{figure*}[htbp]
    \centering
    \includegraphics[width=1.0\linewidth]{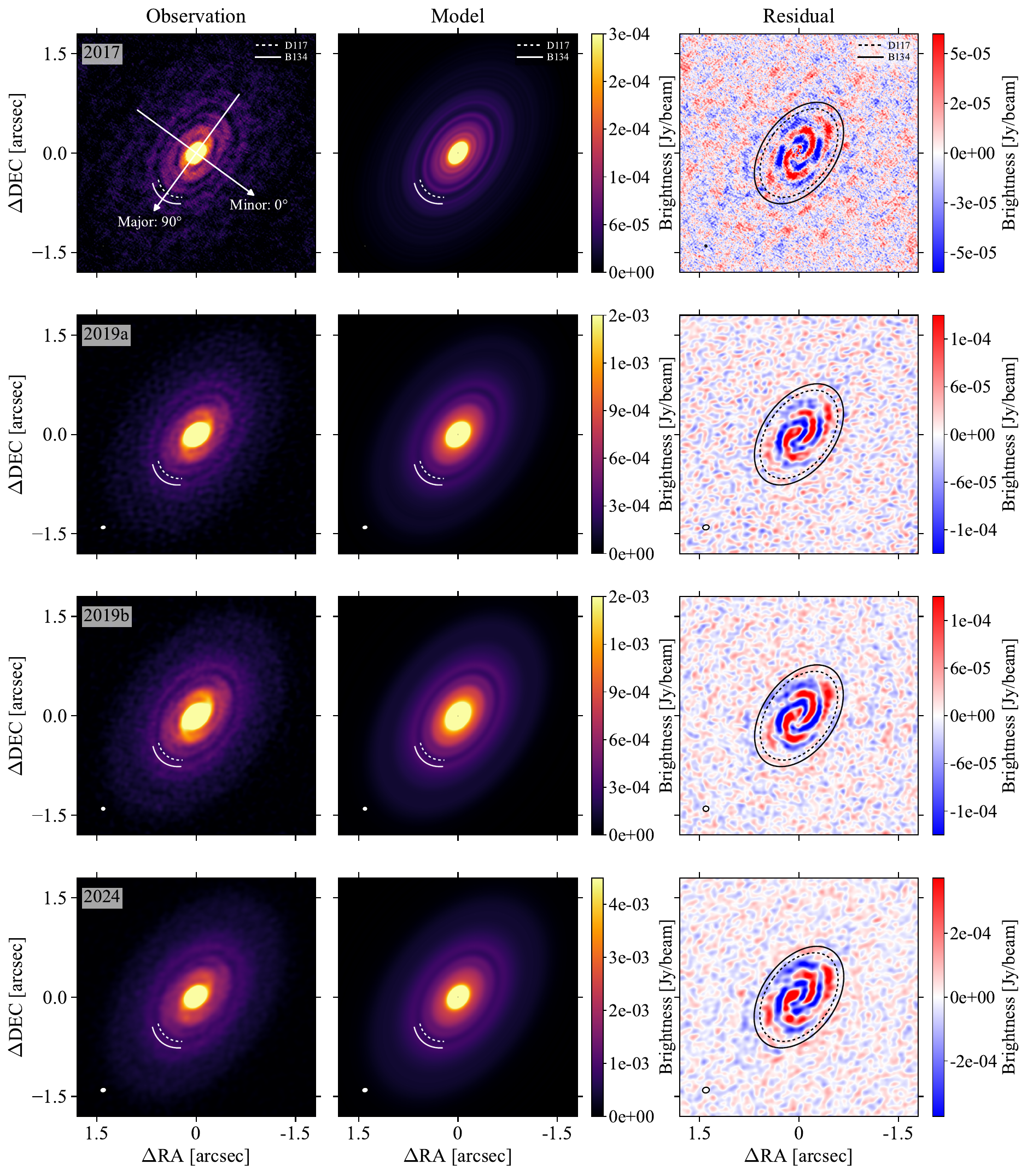}
    \caption{Observed frame images of IM~Lup for the four ALMA epochs. The left, middle, and right columns show the observed CLEAN images, the best fit \texttt{protomidpy} axisymmetric models convolved with CLEAN beam, and the residual images after model subtraction, respectively. The white dotted and solid curves mark the D117 gap and B134 ring from \citet{Huang2018a}, and the white ellipses show the synthesized beams. The white arrows in the observed CLEAN images indicate the major- and minor-axis directions following the coordinate convention adopted in \texttt{protomidpy}, in which north corresponds to ${\rm PA}=0^\circ$ in the observed frame. These directions correspond to the major and minor axes shown in the deprojected images in the following figures.}
    \label{fig:obs_model_residual}
\end{figure*}
\begin{figure*}
    \centering
    \includegraphics[width=1.0\linewidth]{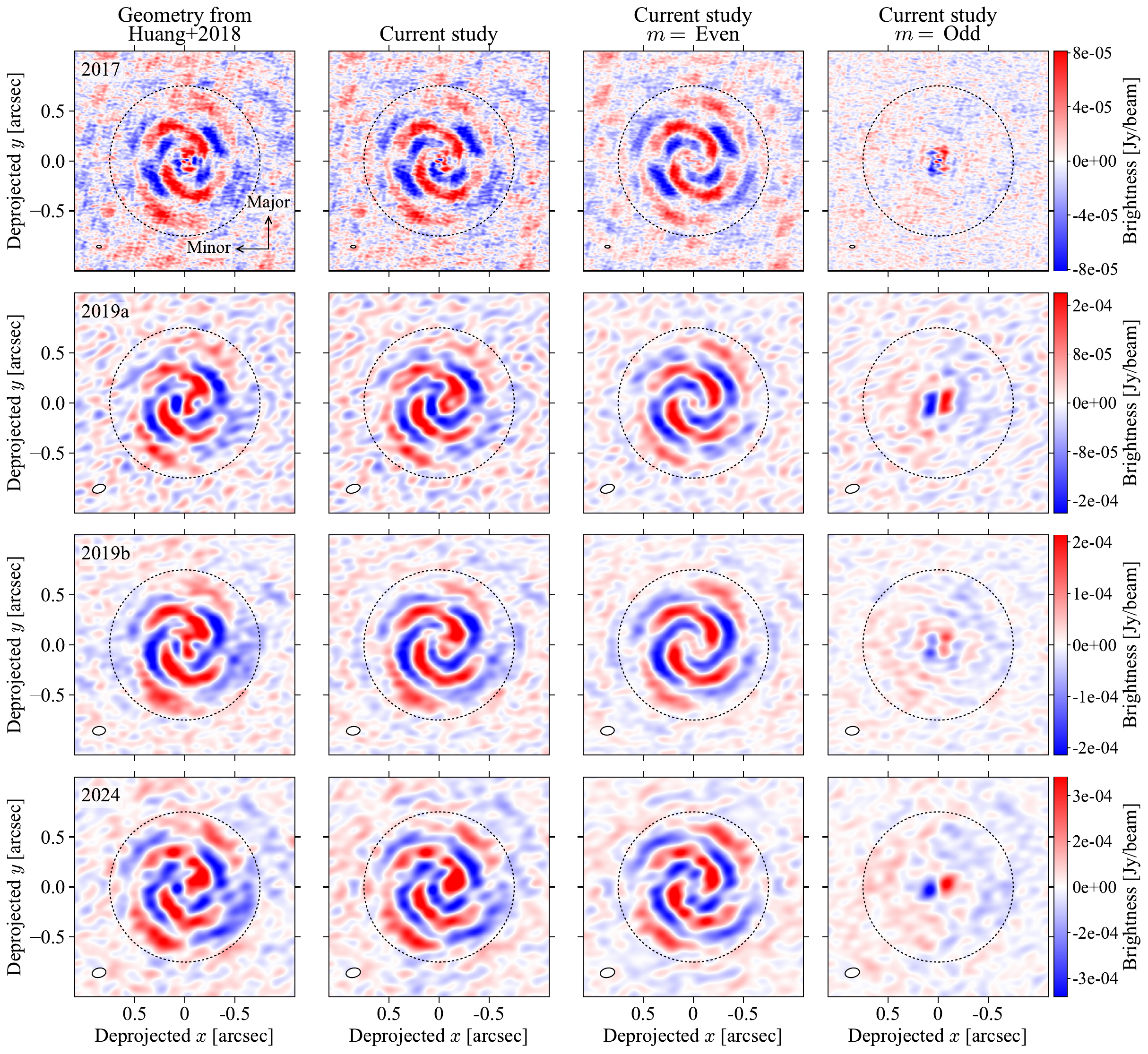}
    \caption{Deprojected frame residual images of IM~Lup for the four epochs. From left to right, the columns show the residual images obtained using the geometry from \citet{Huang2018a}, the residual images obtained using the \texttt{protomidpy} geometry, and their even- and odd-symmetric components. The even- and odd-symmetric residual images are constructed as described in Section~\ref{subsec:residual_imaging}. The black dotted circle marks the D117 gap from \citet{Huang2018a}, and the white ellipses show the synthesized beams.
}
    \label{fig:deprojected_residuals}
\end{figure*}
\begin{figure*}
    \centering
    \includegraphics[width=1.0\linewidth]{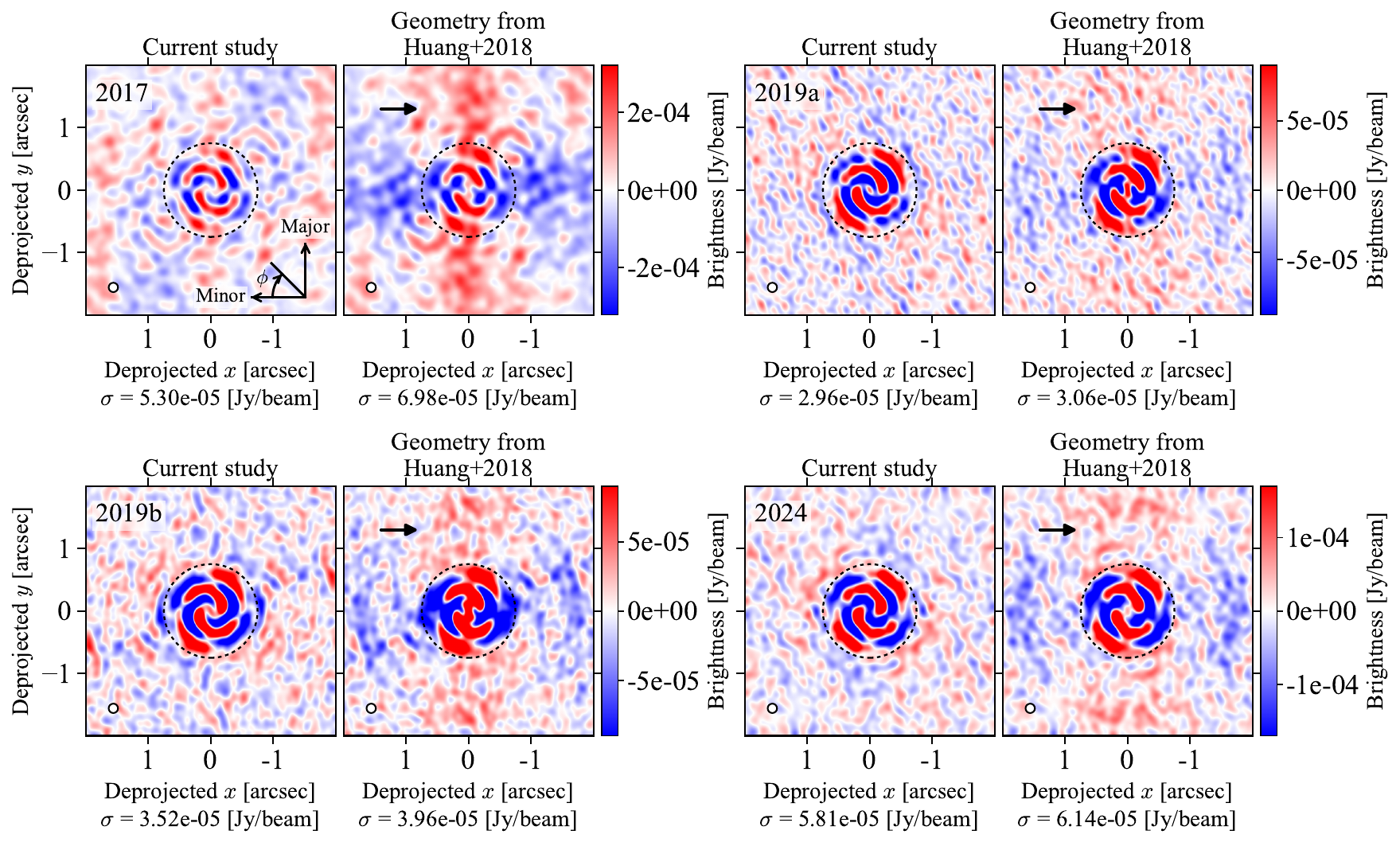}
    \caption{Comparison of deprojected even-symmetric residual images constructed with the \texttt{protomidpy} geometry and with the geometry derived by \citet{Huang2018a}. For each epoch, the left and right panels show the residual images obtained with the \texttt{protomidpy} and \citet{Huang2018a} geometries, respectively. These images are convolved with a common beam of 150 mas and the white ellipses show the synthesized beams. The black dotted circle marks the D117 gap from \citet{Huang2018a}.
    For the geometry from \citet{Huang2018a}, a coherent residual structure aligned approximately with the disk major axis is consistently seen beyond $\sim120$~au in all epochs, as indicated by the arrows. The curved arrow labeled $\phi$ indicates the direction of increasing azimuthal angle used to define the Fourier phase in the following analysis.}
    \label{fig:geometry_effect}
\end{figure*}

Figure~\ref{fig:obs_model_residual} shows the observed CLEAN images, best fit axisymmetric disk models convolved with CLEAN beam, and residual images in the observed frame for the four epochs.
Spiral structures are recovered in the residual images at all epochs, and their overall morphology is broadly similar across the different observation.

\subsubsection{Deprojected residual}

Figure~\ref{fig:deprojected_residuals} shows the deprojected residual images for each epoch. From left to right, the panels show the residual image obtained using the geometry from \citet{Huang2018a}, that obtained using the geometry inferred in this study, and the even- and odd- symmetric components constructed from the real and imaginary parts of the residual visibilities, respectively.
Comparison of the two leftmost panels shows that the residual morphology depends noticeably on the adopted disk geometry, particularly in the inner region around $\sim20$~au.
The residual structure obtained with the geometry from \citet{Huang2018a} also appears more squeezed along the projected minor-axis direction than that obtained with the geometry inferred in this study.
Across all epochs, the outer arm in the northwest region of the observed frame (Figure~\ref{fig:obs_model_residual}), corresponding to the $y<0$ side in the deprojected images of Figure~\ref{fig:deprojected_residuals}, is consistently brighter than the opposite arm.
This arm lies close to the location of the previously reported velocity kink \citep{Pinte2020} and the gap, suggesting the possible presence of an embedded perturbing source.

The even-symmetric residual images clearly recover the $m=2$ spiral pattern at all epochs.
The spiral arms extend roughly from 20 to 100~au, although the residual intensity becomes weaker around $r \sim 80$~au.
The origin of this local minimum is unclear.
In the central region, the Band~6 epochs (2017 and 2019) images show a bar-like structure similar to that discussed by \citet{Aizawa2026a}.

The odd-symmetric residual images show a minor-axis-symmetric residual structure around the disk center in all epochs, with an emission excess on the northeastern, far side of the disk, as inferred from the $^{12}$CO emission  \citep{Pinte2018} and scattered light \citep{Avenhaus2018}.
This asymmetry is discussed as a possible geometric viewing effect associated with finite disk thickness in \citep{Huang2018b,Jennings2022b}.
For IM~Lup, vertically extended millimeter sized grains in the inner region au have been suggested from CO isotopologue modeling \citep{Bosman2023}, and \citet{Ueda2024} proposes an accretion heated, vertically puffed-up inner $\lesssim20$ au region in a gravitationally self-regulated disk model.
The central residuals in odd-symmetric residual are qualitatively consistent with these interpretations and may provide continuum-based evidence for vertical structure in the inner disk.
\begin{figure*}[t]
    \centering
    \includegraphics[width=1.0\linewidth]{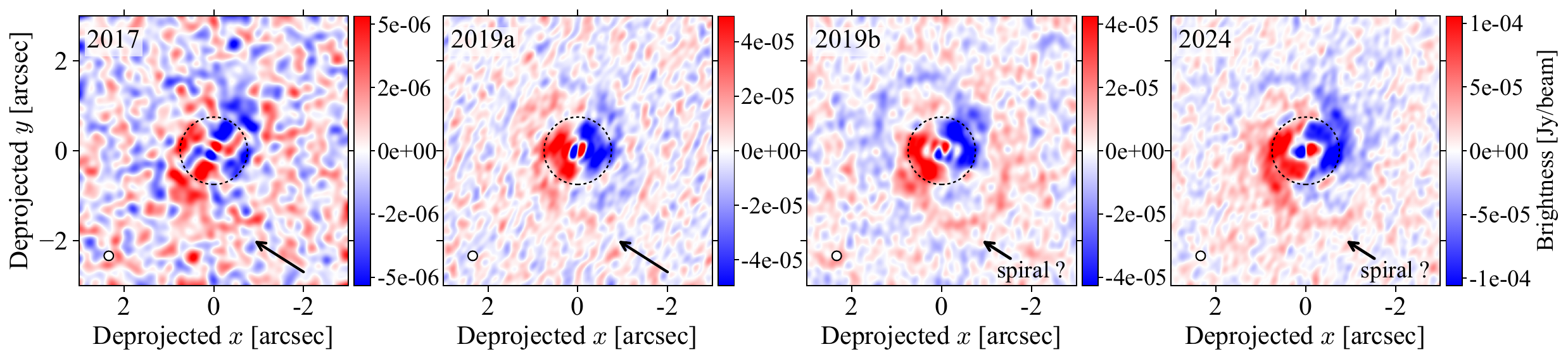}
    \caption{Zoomed-out view of the deprojected odd-symmetric residual images for the four epochs, corresponding to those shown in Figure~\ref{fig:deprojected_residuals}. All images are produced using Briggs weighting with a robust parameter of 0.5 and convolved to a common circular beam of 210~mas. The black dotted circle marks the D117 gap from \citet{Huang2018a}.}
    \label{fig:zoom-out_odd}
\end{figure*}

\subsubsection{Effect of adopted geometry on the residual images}

Figure~\ref{fig:geometry_effect} compares the deprojected even-symmetric residual images used for the pattern speed analysis, all convolved to a common circular beam of 150~mas, using the geometry obtained in the current study and that from \citet{Huang2018a}.
When the \citet{Huang2018a} geometry is used, a coherent quadruple (i.e., $m=2$) residual pattern appears at radii beyond approximately 120~au in all epochs, whereas no similarly common pattern is clearly seen with the geometry with \texttt{protomidpy}.

A mismatch in the adopted inclination is known to produce a quadruple pattern aligned with the disk major axis in the residual map \citep[e.g.,][]{Aizawa2024}, and the pattern seen with the \citet{Huang2018a} geometry closely resembles this characteristic structures. The geometry adopted by \citet{Huang2018a} differs from our estimate, especially in inclination ($\Delta i \approx 3 ^\circ$), and thus, the quadruple residuals are  likely dominated by the effect of the inclination mismatch. In addition, similar major-axis-symmetric residuals appear to extend into the inner disk, including the innermost region, and are especially prominent in the 2017 and 2019 Band~6 data. As shown in Section~\ref{sec4}, these geometry-induced residuals substantially affect the measured pattern speed.
We also checked the odd-symmetric residuals and found no comparable systematic difference between the two geometries.
\begin{figure*}
    \centering
    \includegraphics[width=0.95\linewidth]{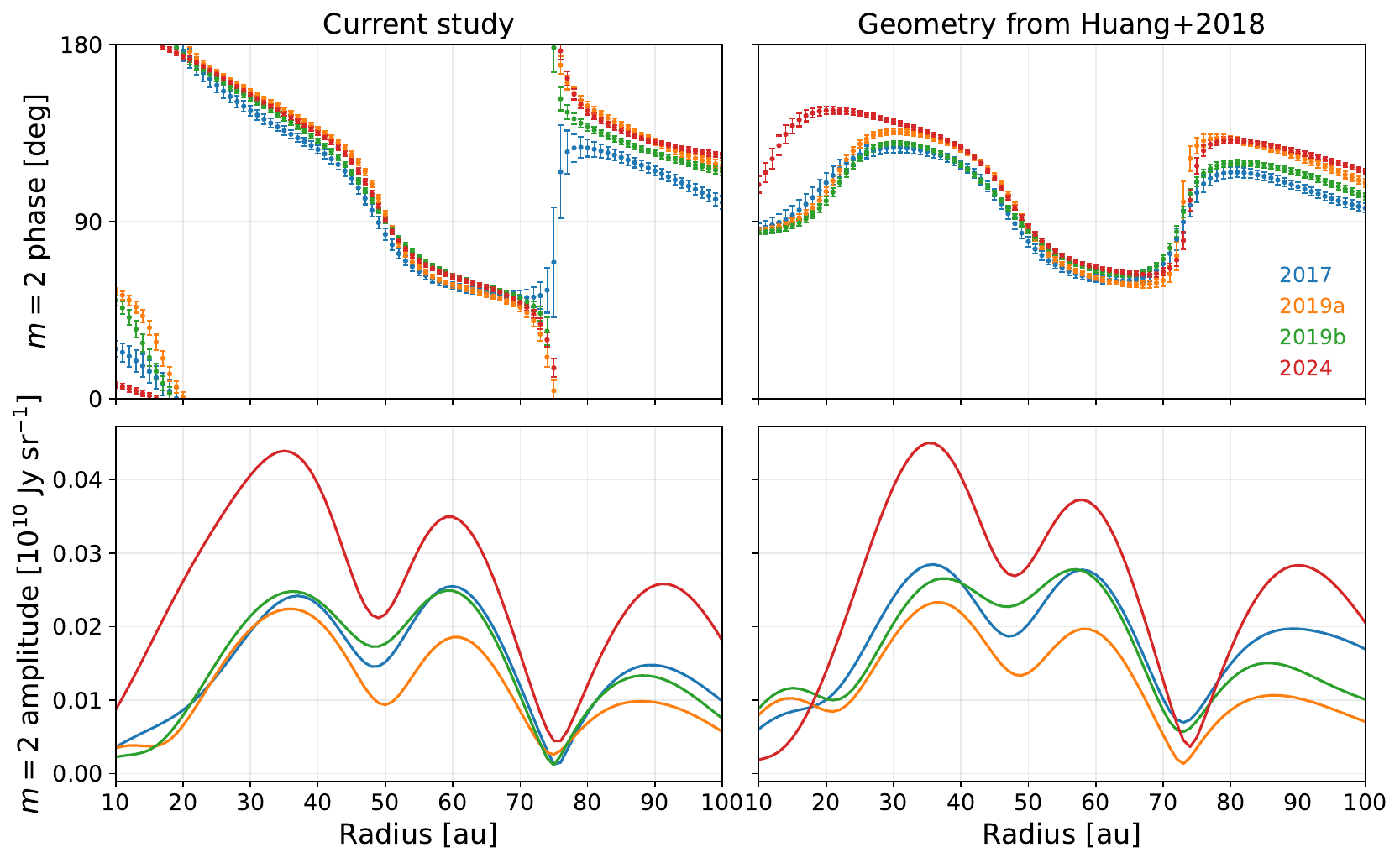}
    \caption{$m=2$ spiral phases and amplitudes obtained from Fourier analysis of the images shown in Figure~\ref{fig:geometry_effect}. The left and right columns show the results obtained using the \texttt{protomidpy} geometry and the geometry from \citet{Huang2018a}, respectively. The upper panels show the $m = 2$ phase, while the lower panels show the corresponding amplitude. Different colors represent the four epochs. All residual images are convolved to a common circular beam of 150~mas before the Fourier analysis. At radii where the $m=2$ amplitude becomes very small, such as around $\sim75$~au, the recovered Fourier phase may not reliably trace the spiral arm.}
    \label{fig:spiral_phase_amp}
\end{figure*}
\subsubsection{Possible one-armed spiral}
Before turning to the pattern speed analysis, we briefly discuss an interesting new feature that is serendipitously identified while testing several imaging parameters. Figure~\ref{fig:zoom-out_odd} presents zoomed-out views of the odd-symmetric residual images produced with a Briggs robust parameter of 0.5 and convolved to a common circular beam of 210~mas, which is larger than the fiducial 150~mas beam and enhances sensitivity to extended structures. In the higher frequency 2019b and 2024 epochs, a large-scale, one-armed spiral-like feature is identified, extending to $\sim 1.5$~arcsec ($\sim 240$ au) and winding in the trailing. Remarkably, the feature appears at approximately the same location in the two epochs, while hints of a similar feature can also be seen in the 2017 and 2019a data. The extended arm is intrinsically faint, with a typical intensity of only about $1$--$2$ times the rms noise measured in the odd-symmetric image, in the 2024 epoch.  
The persistence of the morphology across independent epochs and observing frequencies suggests that the structure may represent genuine non-axisymmetric emission rather than an artifact arising from imaging or geometric misalignment. A similar structure is reported in Elias~27 by \citet{Aizawa2026a}. By decomposing the residual emission into even and odd symmetric components, they identified a possible one-armed spiral in the odd-symmetric residuals, in addition to the well-known $m = 2$ spiral \citep[e.g.,][]{perez2016,Huang2018b}.
A broad minor-axis-symmetric residual structure, with brighter emission on the near side of the disk, is also seen around the base of the possible one-armed spiral. In Appendix~\ref{app:Comparison of the Possible One-Armed Spiral}, we compare this possible one-armed spiral with planetary wake models based on linear density-wave theory.
\section{Spiral Amplitude, Phase and Pattern Speed}
\label{sec4}
As described in Section~\ref{subsec:fourier_analysis}, the spiral morphology and pattern speed are measured from the even-symmetric residual images using Fourier analysis.
\subsection{Spiral amplitude and phase}
\label{subsec:m2_amplitude_phase}
Figure~\ref{fig:spiral_phase_amp} shows the $m=2$ spiral amplitudes and phases obtained from Fourier analysis of the even-symmetric residual images presented in Figure~\ref{fig:geometry_effect}.

The $m=2$ amplitude profiles show similar radial variations across all epochs and for both geometries. Within $\sim100$~au, these variations are consistent with previous studies, including the local minima reported by \citet{Huang2018a}, while \citet{Aizawa2026a} find corresponding changes in the $m=2$ Fourier amplitude and phase associated with rings, gaps, and spiral deflections. However, as is also evident in Figure~\ref{fig:geometry_effect}, the difference between the two geometries becomes substantial beyond $\sim120$~au; for example, in the 2017 epoch, the $m=2$ amplitude obtained with the \citet{Huang2018a} geometry is more than three times larger than that obtained with the \texttt{protomidpy} geometry, demonstrating that the former leaves systematically stronger $m = 2$ residuals.
\begin{figure*}
    \centering
    \includegraphics[width=0.95\linewidth]{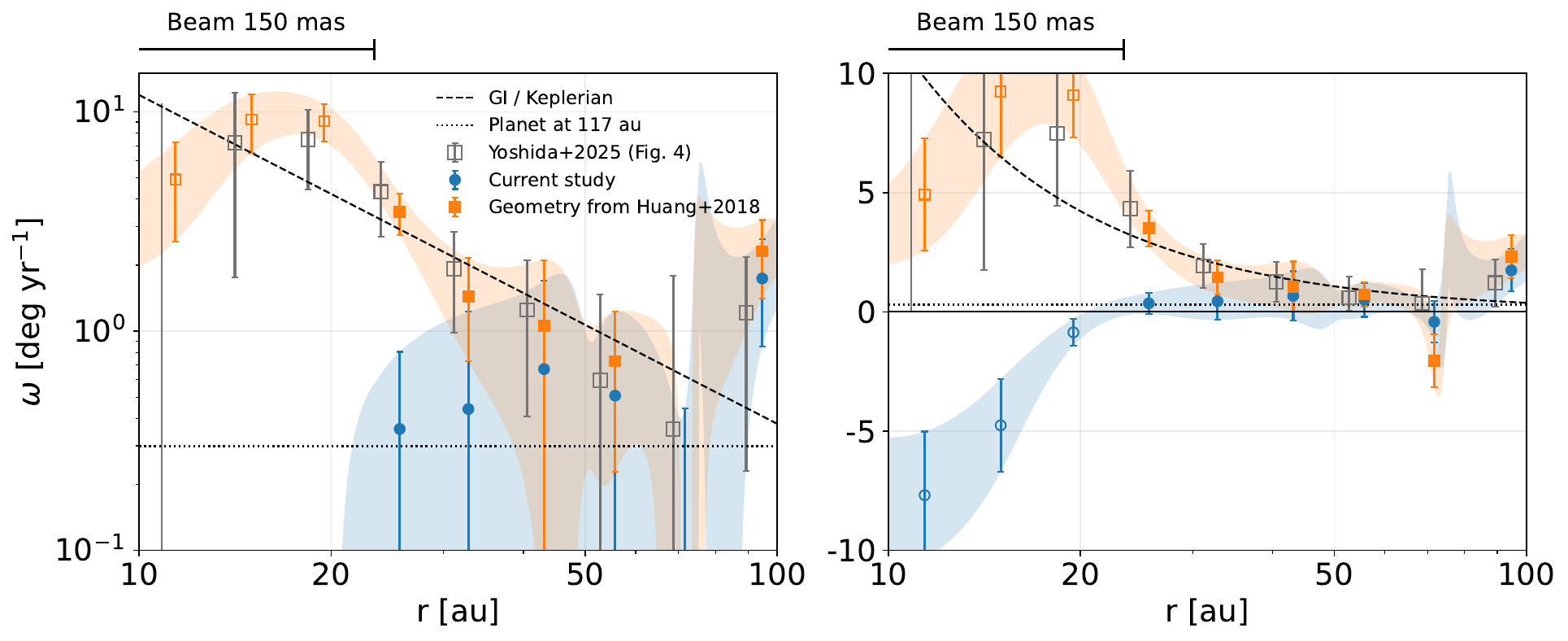}

    \vspace{0.1cm}

    \includegraphics[width=0.95\linewidth]{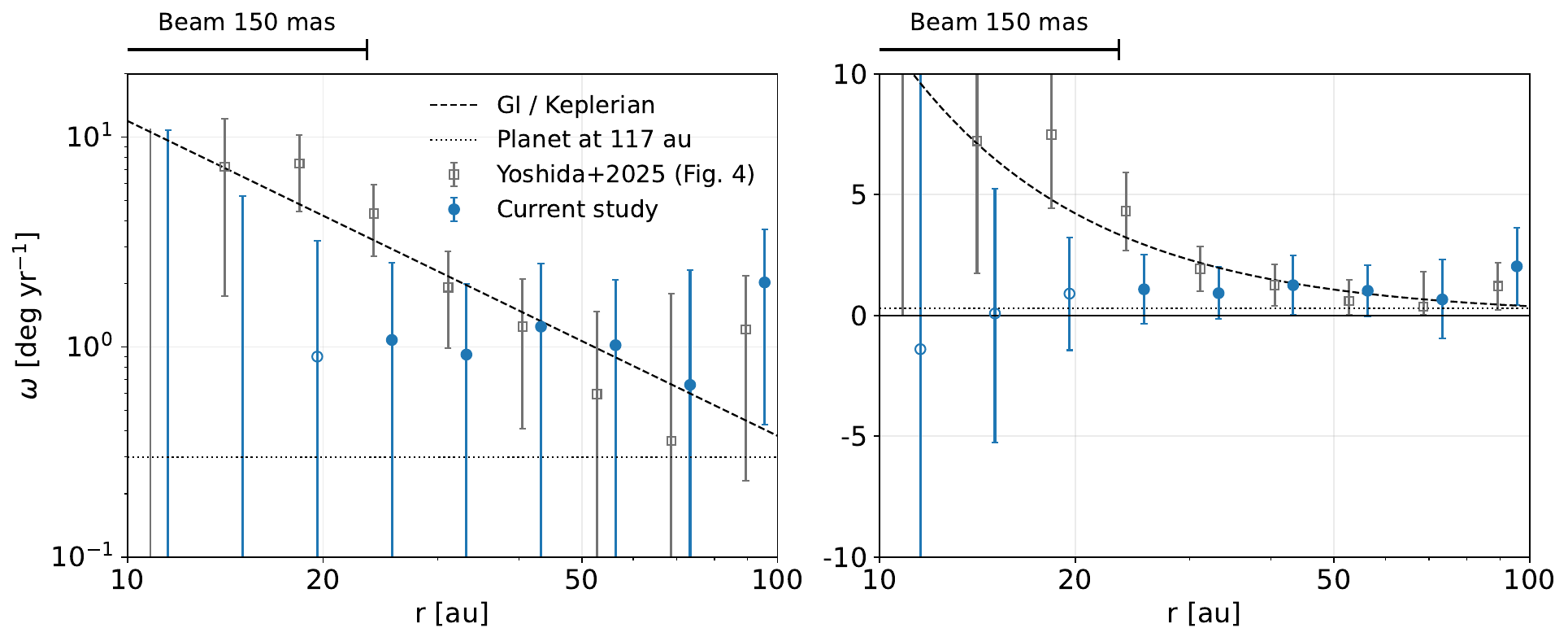}
    \caption{
    Pattern speed profiles of the IM~Lup spiral obtained using two analysis methods.
    The upper panels show the results of our Fourier analysis of the deprojected even-symmetric residual images.
    The blue circles and orange squares indicate the measurements obtained with the \texttt{protomidpy} and \citet{Huang2018a} geometries, respectively. The shaded regions show the uncertainty ranges of the inferred pattern speeds.
    The lower panels show our reanalysis of the Radius-by-radius method of \citet{Yoshida2025} using the \texttt{protomidpy} geometry, with the $m=1$ component excluded, compared with the result of \citet{Yoshida2025}.
    The black dashed and dotted lines indicate the local Keplerian angular velocity and the Keplerian angular velocity at 117~au, respectively.
    The horizontal bars at the upper left of each panel indicate the radial scale corresponding to the common 150~mas beam ($\sim24$~au) measured from the disk center. Open blue and orange markers denote measurements within this scale. The gray open squares show the results reported by \citet{Yoshida2025}.
    The left panels use logarithmic scales, while the right panels use a linear vertical scale.
    }
    \label{fig:pattern_speed}
\end{figure*}

We also find clear differences in the spiral phase profiles.
In particular, with the \citet{Huang2018a} geometry, the radial phase gradient $d\theta/dr$ reverses sign around 20--30~au in all epochs. In contrast, the phase profile obtained with the \texttt{protomidpy} geometry increases monotonically with radius. Because the spiral arms in the IM~Lup disk are interpreted as trailing \citep{Huang2018b}, the phase reversal seen only with the \citet{Huang2018a} geometry suggests that the adopted geometry distorts the apparent spiral morphology.

Toward the innermost region, the phases obtained with the \citet{Huang2018a} geometry converge to approximately $90^\circ$, a trend that is particularly pronounced in the three Band~6 epochs, as shown in the upper right panel of Figure~\ref{fig:spiral_phase_amp}. As shown in Figure~\ref{fig:geometry_effect}, an inclination mismatch can induce a coherent quadrupolar residual pattern aligned with the disk major axis. The measured phases tend to be pulled toward the phase of this geometry-induced component, around $90^\circ$, particularly at radii where the $m=2$ amplitude is relatively weak. This provides a natural explanation for the behavior of $d\theta/dr$ in the inner disk.
\subsection{Spiral pattern speed}
\label{subsec:pattern_speed_result}
Figure~\ref{fig:pattern_speed} shows the pattern speed profiles of the spiral obtained using two methods. The upper panels show the results of our Fourier analysis of the even-symmetric residual images, comparing the geometries inferred with \texttt{protomidpy} and adopted by \citet{Huang2018a}. The lower panels show our reanalysis of the Radius-by-radius method of \citet{Yoshida2025} (Appendix~\ref{app:radiusbyradius}), applied to the deprojected images without axisymmetric model subtraction using the geometry inferred in this study. We note that the pattern speed profiles are correlated on the scale of the beam ($\sim$24 au). Moreover, within one beam width of the disk center, the uncertainty estimation described in Appendix \ref{app:Uncertainty} may underestimate the statistical uncertainties, because each annulus contains only a few independent beams.

In both methods, the pattern speeds obtained with the \citet{Huang2018a} geometry approximately follow the local Keplerian rotation profile, consistent with the results of \citet{Yoshida2025}. In contrast, with the \texttt{protomidpy} geometry, the inferred pattern speeds are consistent with zero within $\sim2\sigma$ at all radii beyond approximately 20~au (the innermost radius lies just inside the beam scale), although the Fourier analysis suggests a marginal positive deviation at the outermost radius (95~au). Most of the measured pattern speeds are positive, hinting at slow rotation of the spiral; however, this may not be statistically significant given the potentially underestimated uncertainties and the correlations between neighboring measurements. The Fourier analysis constrains the pattern speed to $\omega \lesssim 1$--$2$~deg~yr$^{-1}$. The inferred range encompasses the Keplerian angular velocity at $\sim117$~au ($\sim0.3$~deg~yr$^{-1}$), where the presence of an embedded planet has been discussed \citep[e.g.,][]{Pinte2020}.

At $r \leq 15$~au, the pattern speeds derived from the Fourier analysis with the \texttt{protomidpy} geometry become strongly negative, indicating apparent rotation opposite to the direction of disk rotation. We consider these values less reliable for two reasons. First, these radii are not sufficiently resolved by the current beam, and the uncertainties may be underestimated; indeed, the Radius-by-radius analysis yields results consistent with zero rotation. Second, the 2024 Band~7 data do not clearly recover the same structure seen in the Band~6 data. These radii correspond to bar-like structures first reported in the 2017 data \citep{Aizawa2026a}, which are also recovered in the other Band~6 data but not clearly in the Band~7 data (Figure~\ref{fig:deprojected_residuals}) . The pattern speed measurements at these radii are therefore likely affected by differences in the structures traced between the data sets, even with the updated disk geometry. In Appendix~\ref{app:linear_regression}, we present examples of the pattern speed measurements at these radii. Future high-resolution Band~6 observations will be essential for refining the pattern speed measurements in both the inner and outer disks.
\section{Discussion}
\label{sec5}

\subsection{Summary of analysis}
\label{subsec:Summary of analysis}
\begin{figure}
    \centering
    \includegraphics[width=1.0\linewidth]{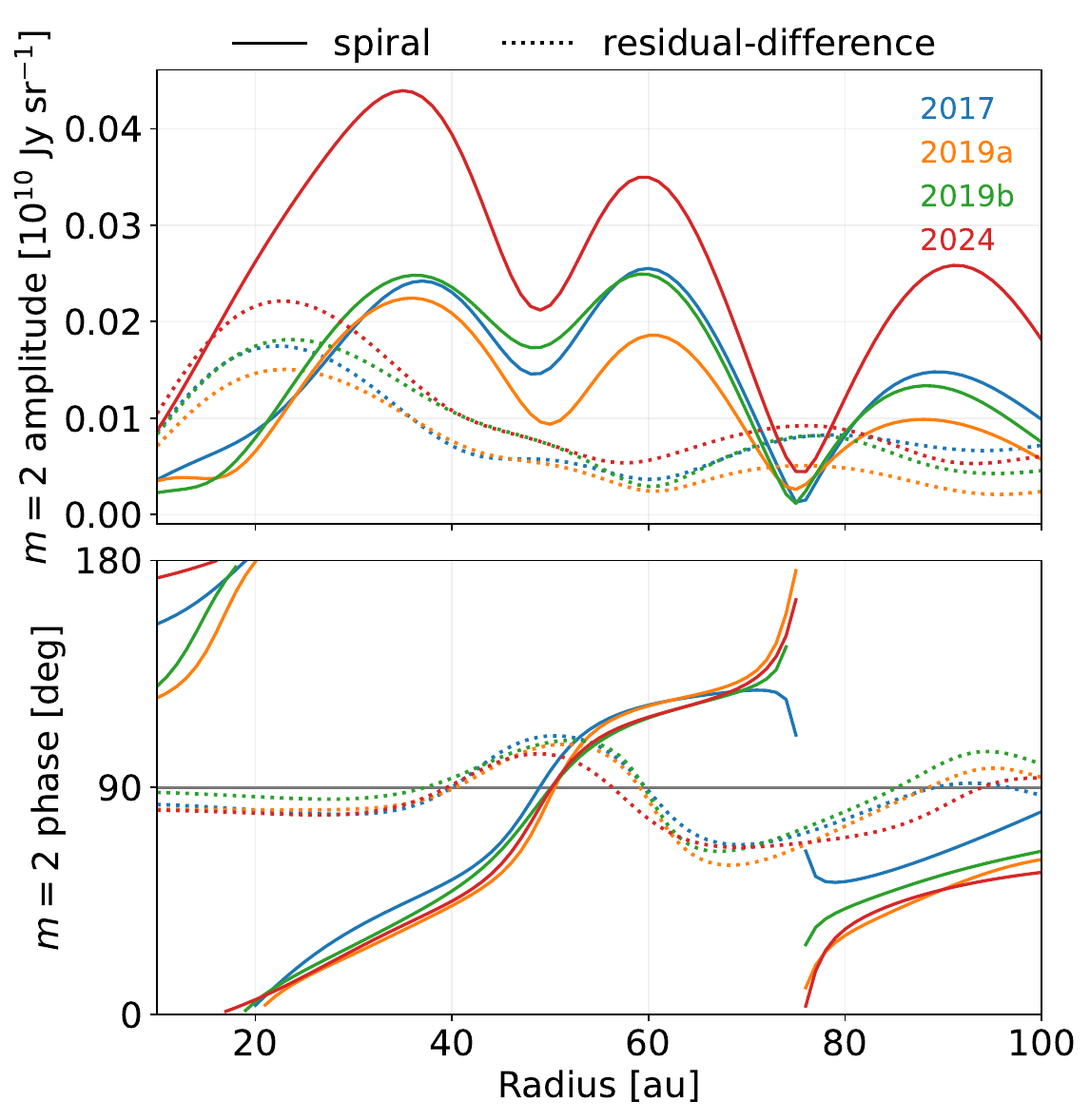}
    \caption{Comparison of the $m = 2$ phase and amplitude of the residuals used the geometry with \texttt{protomidpy} and the residual-difference components .The upper and lower panels show the $m = 2$ phase and amplitude, respectively. Dot lines represent residual-difference components.}
    \label{fig:residual-difference}
\end{figure}
Our analysis yields a spiral pattern speed different from the Keplerian-like profile reported by \citet{Yoshida2025}. Here, we argue that this discrepancy originates from the adopted disk geometry, and we explain why the Keplerian-like profile is obtained when the \citet{Huang2018a} geometry is adopted.

We first note that multiple lines of evidence favor the geometry inferred in this study and \citet{Aizawa2026a}. First, it leaves smaller residuals in the residual images. This is expected, since our geometry is optimized directly in the visibility plane, while adopting the \citet{Huang2018a} geometry leaves a spurious $m=2$ signal aligned with the disk major axis (Figure~\ref{fig:geometry_effect}). Second, our geometry is consistent with that independently derived from CN emission \citep{Paneque2024}. Third, the $m=2$ phases recover with \citet{Huang2018a} geometry switch between trailing and leading configurations and are pulled toward $90^{\circ}$, along the disk major axis. This is the characteristic signature of an inclination mismatch \citep[e.g.,][]{Aizawa2024} rather than a physical reason.

Then, why does the analysis of \citet{Yoshida2025} recover a near-Keplerian pattern speed? If the measured phases were dominated entirely by the geometric mismatch, all of them would be pulled toward $90^{\circ}$ and no coherent pattern speed emerged. This is not what happens: the 2024 Band~7 data are less affected than the other Band~6 data. To quantify this, Figure \ref{fig:residual-difference} shows the amplitude and phase of the $m=2$ component of the difference between the residual images obtained with the two geometries (the residual-difference component), together with the amplitude and phase of the spiral recovered with our geometry. 
The $m=2$ phase of the residual-difference component remains close to $90^{\circ}$ over a broad radial range, as shown in the lower panel of Figure~\ref{fig:residual-difference}. Because the intrinsic non-axisymmetric disk structure is largely canceled when taking the difference between the two residual images, this component primarily isolates the systematic residual introduced by the difference in the adopted disk geometry, particularly the inclination. Its phase near $90^{\circ}$ is therefore consistent with the characteristic major-axis-aligned $m=2$ residual expected from an inclination mismatch.

In the Band~6 data, the residual-difference component is stronger than the intrinsic spiral signal and becomes dominant at inner region, which is why the measured phases are pulled toward $90^{\circ}$ (Figure~\ref{fig:residual-difference}). In the Band~7 data, by contrast, the spiral itself is relatively strong, so the measured phases are only biased toward $90^{\circ}$ rather than being dominated by the geometry-induced component.

Taken together, when the \citet{Huang2018a} geometry is adopted, the Band~6 phases are more strongly distorted than the 2024 Band~7 phase, increasing the apparent inter-epoch phase difference in the inner disk to mimic local Keplerian motion.
We therefore conclude that Keplerian motion of \citet{Yoshida2025} is an artifact of the adopted geometry, which highlight the critical importance of accurately determining the disk geometry for robust spiral pattern speed measurements.

In Appendix~\ref{app:Reanalysis}, we reproduce Figure~3 of \citet{Yoshida2025} using the two disk geometries to examine how the adopted geometry affects the apparent rotation of the inner disk structure.

\subsection{Possibility of infall-driven spiral}
Our analysis suggests that the spiral pattern may be consistent with slow, nearly rigid rotation or no statistically significant rotation. This result reopens the possibility of a planet-driven origin for the spiral. However, it does not necessarily exclude GI, since the tightly wound morphology of the spiral and the presence of locally coherent Fourier components can still be compatible with gravito-turbulent spirals despite their overall tendency toward local corotation \citep{Bethune2021}.

An alternative possibility is GI modified or sustained by late-stage infall. As discussed by \citet{Yoshida2025}, the disk mass of IM~Lup, $M_{\rm d}\sim0.1\,M_{\star}$ \citep{Teague2021,Lodato2023,Martire2024}, may be too low for an isolated disk to naturally sustain a dominant $m=2$ spiral \citep{Bethune2021}, and such GI-driven spiral may also be short-lives \citep{Hall2019}. The continued infall could alleviate both issues by promoting low-order modes and recurrently regenerating grand design spirals \citep[e.g.,][]{Harsono2011,Tomida2017}. 

Recent simulations suggest that late infall can produce spiral patterns with substantially lower and less Keplerian pattern speeds than expected for isolated GI \citep[e.g.,][]{Longarini2025,Calcino2025,Huhn2026}.
This behavior can be qualitatively consistent with the lower pattern speed profile in this study.

Observationally, interactions between protoplanetary disks and surrounding material are identified or proposed in several relatively evolved systems \citep[e.g.,][]{Huang2020,Huang2021,Speedie2025,GallowaySprietsma2026}.
For IM~Lup, continued environmental accretion is also discussed based on the correlated accretion properties in the Lupus region \citep{Winter2024} and its extended gas \citep[e.g.,][]{Cleeves2016} and dust halo \citep{Das2026}.
The tentative large scale $m=1$ spiral identified in this study may therefore represent a remnant signatures of such infall or external perturbations. 
However, no clear evidence for infall has yet been found, and further analysis is required to test this possibility.
\section{Summary and Conclusion}
\label{sec6}
We apply the axisymmetric modeling framework \texttt{protomidpy} to the same multi-epoch ALMA datasets of IM~Lup used in the previous pattern speed study, simultaneously constraining the disk geometry and axisymmetric intensity distribution. Residual images are then constructed by subtracting the best fit axisymmetric models from the observations. Using these updated geometries and residuals, we remeasure the spiral pattern speed of the IM~Lup disk. Our main results are summarized as follows.

\begin{itemize}
    \item The disk geometric parameters inferred with \texttt{protomidpy} are consistent among the four epochs. Compared with the DSHARP image-based geometry, differences of up to approximately $3^\circ$ in inclination and $1^\circ$ in PA are inferred.

    \item The residual images reveal a clear and coherent $m=2$ spiral structure at all epochs. In addition, the odd-symmetric residuals show a tentative large-scale $m=1$ spiral-like structure.

    \item Several lines of evidence suggest that the geometry inferred in this study provides a more consistent description of the disk than that of \citet{Huang2018a}, including the weaker residual, its agreement with the independently inferred CN emission geometry \citep{Paneque2024}, and the more monotonic spiral phase profile.

    \item The pattern speed profile obtained with the geometry from \citet{Huang2018a} is consistent with local Keplerian rotation, as in the previous analysis, whereas the pattern speeds remeasured with the \texttt{protomidpy} geometry are consistent with zero rotation within the uncertainties.

    \item The Keplerian rotation reported previously can be attributed to spurious quadruple signal introduced by the \cite{Huang2018a} geometry which distorts the spiral phases in the inner disk at Band 6 epochs (2017, 2019) while leaving the stronger Band 7 (2024) spiral less affected. This inflates the phase difference between two epochs so that the inferred pattern speed coincidentally matches the local Keplerian rotation. 
\end{itemize}

This study establishes a robust framework for measuring the pattern speed of faint spirals embedded in disk emission while also highlighting the importance of accurately constraining the disk geometry. For IM~Lup, future DSHARP-quality, high-resolution Band~6 observations, combined with a longer time baseline, should allow the spiral structure to be traced more reliably into the inner disk and thereby provide stronger constraints on the pattern speed.

\section*{acknowledgments}
We thank Tomohiro C. Yoshida for providing the the calibrated data and for valuable discussions.
This work was supported by JSPS KAKENHI grant No. 25K17431. Data analysis was in part carried out on the Multi-wavelength Data Analysis System operated by the AstronomyData Center (ADC), National Astronomical Observatory of Japan. This paper makes use of the following ALMA data: 2013.1.00226.S, 2013.1.00694.S, 2013.1.00798.S, 2016.1.00484.L, 2018.1.01055.L, and 2023.1.00525.S. ALMA is a partnership of ESO (representing its member states), NSF (USA) and NINS (Japan), together with NRC (Canada), NSTC and ASIAA (Taiwan), and KASI (Republic of Korea), in cooperation with the Republic of Chile. The Joint ALMA Observatory is operated by ESO, AUI/NRAO and NAOJ. We acknowledge the use of ChatGPT (OpenAI) and Claude (Anthropic) for grammatical and clarity improvements throughout the manuscript.

\section*{Data Availability}
The 2017 and 2019 data analyzed in this article are available in the DSHARP Data Release at \url{https://bulk.cv.nrao.edu/almadata/lp/DSHARP} and the MAPS Data Release at 
\url{https://alma-maps.info/data.html}, respectively. 
The 2024 data are from project code 2023.1.00525.S and are publicly available in the ALMA archive.

{\it Software}: {\tt Astropy} \citep{Astropy2013}, {\tt CASA} \citep{Mcmullin2007}, {\tt corner} \citep{Foreman2016}, {\tt emcee} \citep{Foreman2013}, {\tt Jupyter Notebook} \citep{Kluyver2016}, {\tt Matplotlib} \citep{Hunter2007}, {\tt NumPy} \citep{Walt2011}, {\tt Pandas} \citep{Mckinney-proc-scipy-2010}, {\tt protomidpy} \citep{Aizawa2024}, {\tt SciPy}  \citep{virtanen2020}

\bibliographystyle{apj}
\bibliography{ref}

@ARTICLE{Aizawa2024,
       author = {{Aizawa}, Masataka and {Muto}, Takayuki and {Momose}, Munetake},
        title = "{Revealing asymmetry on mid-plane of protoplanetary disc through modelling of axisymmetric emission: methodology}",
      journal = {\mnras},
         year = 2024,
        month = aug,
       volume = {532},
       number = {2},
        pages = {1361-1390},
          doi = {10.1093/mnras/stae1549},
archivePrefix = {arXiv},
       eprint = {2406.14006},
 primaryClass = {astro-ph.EP},
       adsurl = {https://ui.adsabs.harvard.edu/abs/2024MNRAS.532.1361A}
}

@ARTICLE{Aizawa2026a,
       author = {{Aizawa}, Masataka and {Muto}, Takayuki and {Momose}, Munetake},
        title = "{Axisymmetric Modeling of DSHARP Dusty Disks: Asymmetric Structures and Inner-disk Dispersal}",
      journal = {\apj},
         year = 2026,
        month = feb,
       volume = {998},
       number = {2},
          eid = {225},
        pages = {225},
          doi = {10.3847/1538-4357/ae2fb2},
archivePrefix = {arXiv},
       eprint = {2512.16091},
 primaryClass = {astro-ph.EP},
       adsurl = {https://ui.adsabs.harvard.edu/abs/2026ApJ...998..225A}
}

@ARTICLE{Andrews2018,
       author = {{Andrews}, Sean M. and {Huang}, Jane and {P{\'e}rez}, Laura M. and {Isella}, Andrea and {Dullemond}, Cornelis P. and {Kurtovic}, Nicol{\'a}s T. and {Guzm{\'a}n}, Viviana V. and {Carpenter}, John M. and {Wilner}, David J. and {Zhang}, Shangjia and {Zhu}, Zhaohuan and {Birnstiel}, Tilman and {Bai}, Xue-Ning and {Benisty}, Myriam and {Hughes}, A. Meredith and {{\"O}berg}, Karin I. and {Ricci}, Luca},
        title = "{The Disk Substructures at High Angular Resolution Project (DSHARP). I. Motivation, Sample, Calibration, and Overview}",
      journal = {\apjl},
         year = 2018,
        month = dec,
       volume = {869},
       number = {2},
          eid = {L41},
        pages = {L41},
          doi = {10.3847/2041-8213/aaf741},
archivePrefix = {arXiv},
       eprint = {1812.04040},
 primaryClass = {astro-ph.SR},
       adsurl = {https://ui.adsabs.harvard.edu/abs/2018ApJ...869L..41A}
}

@ARTICLE{Andrews2021,
       author = {{Andrews}, Sean M. and {Elder}, William and {Zhang}, Shangjia and {Huang}, Jane and {Benisty}, Myriam and {Kurtovic}, Nicol{\'a}s T. and {Wilner}, David J. and {Zhu}, Zhaohuan and {Carpenter}, John M. and {P{\'e}rez}, Laura M. and {Teague}, Richard and {Isella}, Andrea and {Ricci}, Luca},
        title = "{Limits on Millimeter Continuum Emission from Circumplanetary Material in the DSHARP Disks}",
      journal = {\apj},
         year = 2021,
        month = jul,
       volume = {916},
       number = {1},
          eid = {51},
        pages = {51},
          doi = {10.3847/1538-4357/ac00b9},
archivePrefix = {arXiv},
       eprint = {2105.08821},
 primaryClass = {astro-ph.EP},
       adsurl = {https://ui.adsabs.harvard.edu/abs/2021ApJ...916...51A}
}

@ARTICLE{Astropy2013,
       author = {{Astropy Collaboration} and {Robitaille}, Thomas P. and {Tollerud}, Erik J. and {Greenfield}, Perry and {Droettboom}, Michael and {Bray}, Erik and {Aldcroft}, Tom and {Davis}, Matt and {Ginsburg}, Adam and {Price-Whelan}, Adrian M. and {Kerzendorf}, Wolfgang E. and {Conley}, Alexander and {Crighton}, Neil and {Barbary}, Kyle and {Muna}, Demitri and {Ferguson}, Henry and {Grollier}, Fr{\'e}d{\'e}ric and {Parikh}, Madhura M. and {Nair}, Prasanth H. and {Unther}, Hans M. and {Deil}, Christoph and {Woillez}, Julien and {Conseil}, Simon and {Kramer}, Roban and {Turner}, James E.~H. and {Singer}, Leo and {Fox}, Ryan and {Weaver}, Benjamin A. and {Zabalza}, Victor and {Edwards}, Zachary I. and {Azalee Bostroem}, K. and {Burke}, D.~J. and {Casey}, Andrew R. and {Crawford}, Steven M. and {Dencheva}, Nadia and {Ely}, Justin and {Jenness}, Tim and {Labrie}, Kathleen and {Lim}, Pey Lian and {Pierfederici}, Francesco and {Pontzen}, Andrew and {Ptak}, Andy and {Refsdal}, Brian and {Servillat}, Mathieu and {Streicher}, Ole},
        title = "{Astropy: A community Python package for astronomy}",
      journal = {\aap},
         year = 2013,
        month = oct,
       volume = {558},
          eid = {A33},
        pages = {A33},
          doi = {10.1051/0004-6361/201322068},
archivePrefix = {arXiv},
       eprint = {1307.6212},
 primaryClass = {astro-ph.IM},
       adsurl = {https://ui.adsabs.harvard.edu/abs/2013A&A...558A..33A}
}

@ARTICLE{Avenhaus2018,
       author = {{Avenhaus}, Henning and {Quanz}, Sascha P. and {Garufi}, Antonio and {Perez}, Sebastian and {Casassus}, Simon and {Pinte}, Christophe and {Bertrang}, Gesa H.-M. and {Caceres}, Claudio and {Benisty}, Myriam and {Dominik}, Carsten},
        title = "{Disks around T Tauri Stars with SPHERE (DARTTS-S). I. SPHERE/IRDIS Polarimetric Imaging of Eight Prominent T Tauri Disks}",
      journal = {\apj},
         year = 2018,
        month = aug,
       volume = {863},
       number = {1},
          eid = {44},
        pages = {44},
          doi = {10.3847/1538-4357/aab846},
archivePrefix = {arXiv},
       eprint = {1803.10882},
 primaryClass = {astro-ph.SR},
       adsurl = {https://ui.adsabs.harvard.edu/abs/2018ApJ...863...44A}
}

@ARTICLE{Bae_Zhu2018,
       author = {{Bae}, Jaehan and {Zhu}, Zhaohuan},
        title = "{Planet-driven Spiral Arms in Protoplanetary Disks. I. Formation Mechanism}",
      journal = {\apj},
         year = 2018,
        month = jun,
       volume = {859},
       number = {2},
          eid = {118},
        pages = {118},
          doi = {10.3847/1538-4357/aabf8c},
archivePrefix = {arXiv},
       eprint = {1711.08161},
 primaryClass = {astro-ph.EP},
       adsurl = {https://ui.adsabs.harvard.edu/abs/2018ApJ...859..118B}
}

@ARTICLE{Bae2018b,
       author = {{Bae}, Jaehan and {Zhu}, Zhaohuan},
        title = "{Planet-driven Spiral Arms in Protoplanetary Disks. II. Implications}",
      journal = {\apj},
         year = 2018,
        month = jun,
       volume = {859},
       number = {2},
          eid = {119},
        pages = {119},
          doi = {10.3847/1538-4357/aabf93},
archivePrefix = {arXiv},
       eprint = {1711.08166},
 primaryClass = {astro-ph.EP},
       adsurl = {https://ui.adsabs.harvard.edu/abs/2018ApJ...859..119B}
}

@ARTICLE{Bethune2021,
       author = {{B{\'e}thune}, William and {Latter}, Henrik and {Kley}, Wilhelm},
        title = "{Spiral structures in gravito-turbulent gaseous disks}",
      journal = {\aap},
         year = 2021,
        month = jun,
       volume = {650},
          eid = {A49},
        pages = {A49},
          doi = {10.1051/0004-6361/202040094},
archivePrefix = {arXiv},
       eprint = {2102.00775},
 primaryClass = {astro-ph.EP},
       adsurl = {https://ui.adsabs.harvard.edu/abs/2021A&A...650A..49B}
}

@ARTICLE{Bosman2023,
       author = {{Bosman}, Arthur D. and {Appelgren}, Johan and {Bergin}, Edwin A. and {Lambrechts}, Michiel and {Johansen}, Anders},
        title = "{A Potential Site for Wide-orbit Giant Planet Formation in the IM Lup Disk}",
      journal = {\apjl},
         year = 2023,
        month = feb,
       volume = {944},
       number = {2},
          eid = {L53},
        pages = {L53},
          doi = {10.3847/2041-8213/acb651},
archivePrefix = {arXiv},
       eprint = {2302.03726},
 primaryClass = {astro-ph.EP},
       adsurl = {https://ui.adsabs.harvard.edu/abs/2023ApJ...944L..53B}
}

@ARTICLE{Calcino2025,
       author = {{Calcino}, Josh and {Price}, Daniel J. and {Hilder}, Thomas and {Christiaens}, Valentin and {Speedie}, Jessica and {Ormel}, Chris W.},
        title = "{Anatomy of a fall: stationary and super-Keplerian spiral arms generated by accretion streamers in protostellar discs}",
      journal = {\mnras},
         year = 2025,
        month = mar,
       volume = {537},
       number = {3},
        pages = {2695-2707},
          doi = {10.1093/mnras/staf135},
archivePrefix = {arXiv},
       eprint = {2410.18521},
 primaryClass = {astro-ph.EP},
       adsurl = {https://ui.adsabs.harvard.edu/abs/2025MNRAS.537.2695C}
}

@ARTICLE{Cleeves2016,
       author = {{Cleeves}, L. Ilsedore and {{\"O}berg}, Karin I. and {Wilner}, David J. and {Huang}, Jane and {Loomis}, Ryan A. and {Andrews}, Sean M. and {Czekala}, Ian},
        title = "{The Coupled Physical Structure of Gas and Dust in the IM Lup Protoplanetary Disk}",
      journal = {\apj},
         year = 2016,
        month = dec,
       volume = {832},
       number = {2},
          eid = {110},
        pages = {110},
          doi = {10.3847/0004-637X/832/2/110},
archivePrefix = {arXiv},
       eprint = {1610.00715},
 primaryClass = {astro-ph.SR},
       adsurl = {https://ui.adsabs.harvard.edu/abs/2016ApJ...832..110C}
}

@ARTICLE{Cossins2009,
       author = {{Cossins}, Peter and {Lodato}, Giuseppe and {Clarke}, C.~J.},
        title = "{Characterizing the gravitational instability in cooling accretion discs}",
      journal = {\mnras},
         year = 2009,
        month = mar,
       volume = {393},
       number = {4},
        pages = {1157-1173},
          doi = {10.1111/j.1365-2966.2008.14275.x},
archivePrefix = {arXiv},
       eprint = {0811.3629},
 primaryClass = {astro-ph},
       adsurl = {https://ui.adsabs.harvard.edu/abs/2009MNRAS.393.1157C}
}

@ARTICLE{Cuello2019,
       author = {{Cuello}, Nicol{\'a}s and {Dipierro}, Giovanni and {Mentiplay}, Daniel and {Price}, Daniel J. and {Pinte}, Christophe and {Cuadra}, Jorge and {Laibe}, Guillaume and {M{\'e}nard}, Fran{\c{c}}ois and {Poblete}, Pedro P. and {Montesinos}, Mat{\'\i}as},
        title = "{Flybys in protoplanetary discs: I. Gas and dust dynamics}",
      journal = {\mnras},
         year = 2019,
        month = mar,
       volume = {483},
       number = {3},
        pages = {4114-4139},
          doi = {10.1093/mnras/sty3325},
archivePrefix = {arXiv},
       eprint = {1812.00961},
 primaryClass = {astro-ph.EP},
       adsurl = {https://ui.adsabs.harvard.edu/abs/2019MNRAS.483.4114C}
}

@ARTICLE{Das2026,
       author = {{Das}, Sreejita and {Mac{\'\i}as}, Enrique and {Kurtovic}, Nicolas T. and {Birnstiel}, Til and {Viscardi}, Elena M. and {Curone}, Pietro},
        title = "{Dust characterization of halos: The extended emission in protoplanetary disks}",
      journal = {\aap},
         year = 2026,
        month = jun,
       volume = {710},
          eid = {A238},
        pages = {A238},
          doi = {10.1051/0004-6361/202659644},
archivePrefix = {arXiv},
       eprint = {2604.18332},
 primaryClass = {astro-ph.EP},
       adsurl = {https://ui.adsabs.harvard.edu/abs/2026A&A...710A.238D}
}

@ARTICLE{Dong2018,
       author = {{Dong}, Ruobing and {Najita}, Joan R. and {Brittain}, Sean},
        title = "{Spiral Arms in Disks: Planets or Gravitational Instability?}",
      journal = {\apj},
         year = 2018,
        month = aug,
       volume = {862},
       number = {2},
          eid = {103},
        pages = {103},
          doi = {10.3847/1538-4357/aaccfc},
archivePrefix = {arXiv},
       eprint = {1806.05183},
 primaryClass = {astro-ph.SR},
       adsurl = {https://ui.adsabs.harvard.edu/abs/2018ApJ...862..103D}
}

@ARTICLE{Foreman2013,
       author = {{Foreman-Mackey}, Daniel and {Hogg}, David W. and {Lang}, Dustin and {Goodman}, Jonathan},
        title = "{emcee: The MCMC Hammer}",
      journal = {\pasp},
         year = 2013,
        month = mar,
       volume = {125},
       number = {925},
        pages = {306},
          doi = {10.1086/670067},
archivePrefix = {arXiv},
       eprint = {1202.3665},
 primaryClass = {astro-ph.IM},
       adsurl = {https://ui.adsabs.harvard.edu/abs/2013PASP..125..306F}
}

@ARTICLE{Foreman2016,
       author = {{Foreman-Mackey}, Daniel},
        title = "{corner.py: Scatterplot matrices in Python}",
      journal = {The Journal of Open Source Software},
         year = 2016,
        month = jun,
       volume = {1},
        pages = {24},
          doi = {10.21105/joss.00024},
       adsurl = {https://ui.adsabs.harvard.edu/abs/2016JOSS....1...24F}
}

@ARTICLE{Forgan2018,
       author = {{Forgan}, Duncan H. and {Ilee}, John D. and {Meru}, Farzana},
        title = "{Are Elias 2-27's Spiral Arms Driven by Self-gravity, or by a Companion? A Comparative Spiral Morphology Study}",
      journal = {\apjl},
         year = 2018,
        month = jun,
       volume = {860},
       number = {1},
          eid = {L5},
        pages = {L5},
          doi = {10.3847/2041-8213/aac7c9},
archivePrefix = {arXiv},
       eprint = {1805.08041},
 primaryClass = {astro-ph.SR},
       adsurl = {https://ui.adsabs.harvard.edu/abs/2018ApJ...860L...5F}
}

@ARTICLE{Fukagawa2004,
       author = {{Fukagawa}, Misato and {Hayashi}, Masahiko and {Tamura}, Motohide and {Itoh}, Yoichi and {Hayashi}, Saeko S. and {Oasa}, Yumiko and {Takeuchi}, Taku and {Morino}, Jun-ichi and {Murakawa}, Koji and {Oya}, Shin and {Yamashita}, Takuya and {Suto}, Hiroshi and {Mayama}, Satoshi and {Naoi}, Takahiro and {Ishii}, Miki and {Pyo}, Tae-Soo and {Nishikawa}, Takayuki and {Takato}, Naruhisa and {Usuda}, Tomonori and {Ando}, Hiroyasu and {Iye}, Masanori and {Miyama}, Shoken M. and {Kaifu}, Norio},
        title = "{Spiral Structure in the Circumstellar Disk around AB Aurigae}",
      journal = {\apjl},
         year = 2004,
        month = apr,
       volume = {605},
       number = {1},
        pages = {L53-L56},
          doi = {10.1086/420699},
       adsurl = {https://ui.adsabs.harvard.edu/abs/2004ApJ...605L..53F}
}

@ARTICLE{Fung2015,
       author = {{Fung}, Jeffrey and {Dong}, Ruobing},
        title = "{Inferring Planet Mass from Spiral Structures in Protoplanetary Disks}",
      journal = {\apjl},
         year = 2015,
        month = dec,
       volume = {815},
       number = {2},
          eid = {L21},
        pages = {L21},
          doi = {10.1088/2041-8205/815/2/L21},
archivePrefix = {arXiv},
       eprint = {1511.01178},
 primaryClass = {astro-ph.EP},
       adsurl = {https://ui.adsabs.harvard.edu/abs/2015ApJ...815L..21F}
}

@ARTICLE{Gaia2023,
       author = {{Gaia Collaboration} and {Vallenari}, A. and {Brown}, A.~G.~A. and {Prusti}, T. and {de Bruijne}, J.~H.~J. and {Arenou}, F. and {Babusiaux}, C. and {Biermann}, M. and {Creevey}, O.~L. and {Ducourant}, C. and {Evans}, D.~W. and {Eyer}, L. and {Guerra}, R. and {Hutton}, A. and {Jordi}, C. and {Klioner}, S.~A. and {Lammers}, U.~L. and {Lindegren}, L. and {Luri}, X. and {Mignard}, F. and {Panem}, C. and {Pourbaix}, D. and {Randich}, S. and {Sartoretti}, P. and {Soubiran}, C. and {Tanga}, P. and {Walton}, N.~A. and {Bailer-Jones}, C.~A.~L. and {Bastian}, U. and {Drimmel}, R. and {Jansen}, F. and {Katz}, D. and {Lattanzi}, M.~G. and {van Leeuwen}, F. and {Bakker}, J. and {Cacciari}, C. and {Casta{\~n}eda}, J. and {De Angeli}, F. and {Fabricius}, C. and {Fouesneau}, M. and {Fr{\'e}mat}, Y. and {Galluccio}, L. and {Guerrier}, A. and {Heiter}, U. and {Masana}, E. and {Messineo}, R. and {Mowlavi}, N. and {Nicolas}, C. and {Nienartowicz}, K. and {Pailler}, F. and {Panuzzo}, P. and {Riclet}, F. and {Roux}, W. and {Seabroke}, G.~M. and {Sordo}, R. and {Th{\'e}venin}, F. and {Gracia-Abril}, G. and {Portell}, J. and {Teyssier}, D. and {Altmann}, M. and {Andrae}, R. and {Audard}, M. and {Bellas-Velidis}, I. and {Benson}, K. and {Berthier}, J. and {Blomme}, R. and {Burgess}, P.~W. and {Busonero}, D. and {Busso}, G. and {C{\'a}novas}, H. and {Carry}, B. and {Cellino}, A. and {Cheek}, N. and {Clementini}, G. and {Damerdji}, Y. and {Davidson}, M. and {de Teodoro}, P. and {Nu{\~n}ez Campos}, M. and {Delchambre}, L. and {Dell'Oro}, A. and {Esquej}, P. and {Fern{\'a}ndez-Hern{\'a}ndez}, J. and {Fraile}, E. and {Garabato}, D. and {Garc{\'\i}a-Lario}, P. and {Gosset}, E. and {Haigron}, R. and {Halbwachs}, J.-L. and {Hambly}, N.~C. and {Harrison}, D.~L. and {Hern{\'a}ndez}, J. and {Hestroffer}, D. and {Hodgkin}, S.~T. and {Holl}, B. and {Jan{\ss}en}, K. and {Jevardat de Fombelle}, G. and {Jordan}, S. and {Krone-Martins}, A. and {Lanzafame}, A.~C. and {L{\"o}ffler}, W. and {Marchal}, O. and {Marrese}, P.~M. and {Moitinho}, A. and {Muinonen}, K. and {Osborne}, P. and {Pancino}, E. and {Pauwels}, T. and {Recio-Blanco}, A. and {Reyl{\'e}}, C. and {Riello}, M. and {Rimoldini}, L. and {Roegiers}, T. and {Rybizki}, J. and {Sarro}, L.~M. and {Siopis}, C. and {Smith}, M. and {Sozzetti}, A. and {Utrilla}, E. and {van Leeuwen}, M. and {Abbas}, U. and {{\'A}brah{\'a}m}, P. and {Abreu Aramburu}, A. and {Aerts}, C. and {Aguado}, J.~J. and {Ajaj}, M. and {Aldea-Montero}, F. and {Altavilla}, G. and {{\'A}lvarez}, M.~A. and {Alves}, J. and {Anders}, F. and {Anderson}, R.~I. and {Anglada Varela}, E. and {Antoja}, T. and {Baines}, D. and {Baker}, S.~G. and {Balaguer-N{\'u}{\~n}ez}, L. and {Balbinot}, E. and {Balog}, Z. and {Barache}, C. and {Barbato}, D. and {Barros}, M. and {Barstow}, M.~A. and {Bartolom{\'e}}, S. and {Bassilana}, J.-L. and {Bauchet}, N. and {Becciani}, U. and {Bellazzini}, M. and {Berihuete}, A. and {Bernet}, M. and {Bertone}, S. and {Bianchi}, L. and {Binnenfeld}, A. and {Blanco-Cuaresma}, S. and {Blazere}, A. and {Boch}, T. and {Bombrun}, A. and {Bossini}, D. and {Bouquillon}, S. and {Bragaglia}, A. and {Bramante}, L. and {Breedt}, E. and {Bressan}, A. and {Brouillet}, N. and {Brugaletta}, E. and {Bucciarelli}, B. and {Burlacu}, A. and {Butkevich}, A.~G. and {Buzzi}, R. and {Caffau}, E. and {Cancelliere}, R. and {Cantat-Gaudin}, T. and {Carballo}, R. and {Carlucci}, T. and {Carnerero}, M.~I. and {Carrasco}, J.~M. and {Casamiquela}, L. and {Castellani}, M. and {Castro-Ginard}, A. and {Chaoul}, L. and {Charlot}, P. and {Chemin}, L. and {Chiaramida}, V. and {Chiavassa}, A. and {Chornay}, N. and {Comoretto}, G. and {Contursi}, G. and {Cooper}, W.~J. and {Cornez}, T. and {Cowell}, S. and {Crifo}, F. and {Cropper}, M. and {Crosta}, M. and {Crowley}, C. and {Dafonte}, C. and {Dapergolas}, A. and {David}, M. and {David}, P. and {de Laverny}, P. and {De Luise}, F. and {De March}, R.},
        title = "{Gaia Data Release 3. Summary of the content and survey properties}",
      journal = {\aap},
         year = 2023,
        month = jun,
       volume = {674},
          eid = {A1},
        pages = {A1},
          doi = {10.1051/0004-6361/202243940},
archivePrefix = {arXiv},
       eprint = {2208.00211},
 primaryClass = {astro-ph.GA},
       adsurl = {https://ui.adsabs.harvard.edu/abs/2023A&A...674A...1G}
}

@ARTICLE{GallowaySprietsma2026,
       author = {{Galloway-Sprietsma}, Maria and {Bae}, Jaehan and {Phillips}, Toni and {Huang}, Jane and {Benisty}, Myriam and {Porter}, Matthew and {Ginski}, Christian and {Winter}, Andrew},
        title = "{A Streamer Driving Misalignment in the Circumtriple Disk of GW Ori}",
      journal = {\aj},
         year = 2026,
        month = sep,
       volume = {172},
       number = {3},
          eid = {138},
        pages = {138},
          doi = {10.3847/1538-3881/ae8bae},
archivePrefix = {arXiv},
       eprint = {2608.06268},
 primaryClass = {astro-ph.EP},
       adsurl = {https://ui.adsabs.harvard.edu/abs/2026AJ....172..138G}
}

@ARTICLE{Grady2013,
       author = {{Grady}, C.~A. and {Muto}, T. and {Hashimoto}, J. and {Fukagawa}, M. and {Currie}, T. and {Biller}, B. and {Thalmann}, C. and {Sitko}, M.~L. and {Russell}, R. and {Wisniewski}, J. and {Dong}, R. and {Kwon}, J. and {Sai}, S. and {Hornbeck}, J. and {Schneider}, G. and {Hines}, D. and {Moro Mart{\'\i}n}, A. and {Feldt}, M. and {Henning}, Th. and {Pott}, J.-U. and {Bonnefoy}, M. and {Bouwman}, J. and {Lacour}, S. and {Mueller}, A. and {Juh{\'a}sz}, A. and {Crida}, A. and {Chauvin}, G. and {Andrews}, S. and {Wilner}, D. and {Kraus}, A. and {Dahm}, S. and {Robitaille}, T. and {Jang-Condell}, H. and {Abe}, L. and {Akiyama}, E. and {Brandner}, W. and {Brandt}, T. and {Carson}, J. and {Egner}, S. and {Follette}, K.~B. and {Goto}, M. and {Guyon}, O. and {Hayano}, Y. and {Hayashi}, M. and {Hayashi}, S. and {Hodapp}, K. and {Ishii}, M. and {Iye}, M. and {Janson}, M. and {Kandori}, R. and {Knapp}, G. and {Kudo}, T. and {Kusakabe}, N. and {Kuzuhara}, M. and {Mayama}, S. and {McElwain}, M. and {Matsuo}, T. and {Miyama}, S. and {Morino}, J.-I. and {Nishimura}, T. and {Pyo}, T.-S. and {Serabyn}, G. and {Suto}, H. and {Suzuki}, R. and {Takami}, M. and {Takato}, N. and {Terada}, H. and {Tomono}, D. and {Turner}, E. and {Watanabe}, M. and {Yamada}, T. and {Takami}, H. and {Usuda}, T. and {Tamura}, M.},
        title = "{Spiral Arms in the Asymmetrically Illuminated Disk of MWC 758 and Constraints on Giant Planets}",
      journal = {\apj},
         year = 2013,
        month = jan,
       volume = {762},
       number = {1},
          eid = {48},
        pages = {48},
          doi = {10.1088/0004-637X/762/1/48},
archivePrefix = {arXiv},
       eprint = {1212.1466},
 primaryClass = {astro-ph.SR},
       adsurl = {https://ui.adsabs.harvard.edu/abs/2013ApJ...762...48G}
}

@ARTICLE{Hall2019,
       author = {{Hall}, Cassandra and {Dong}, Ruobing and {Rice}, Ken and {Harries}, Tim J. and {Najita}, Joan and {Alexander}, Richard and {Brittain}, Sean},
        title = "{The Temporal Requirements of Directly Observing Self-gravitating Spiral Waves in Protoplanetary Disks with ALMA}",
      journal = {\apj},
         year = 2019,
        month = feb,
       volume = {871},
       number = {2},
          eid = {228},
        pages = {228},
          doi = {10.3847/1538-4357/aafac2},
archivePrefix = {arXiv},
       eprint = {1901.02407},
 primaryClass = {astro-ph.EP},
       adsurl = {https://ui.adsabs.harvard.edu/abs/2019ApJ...871..228H}
}

@ARTICLE{Hall2020,
       author = {{Hall}, C. and {Dong}, R. and {Teague}, R. and {Terry}, J. and {Pinte}, C. and {Paneque-Carre{\~n}o}, T. and {Veronesi}, B. and {Alexander}, R.~D. and {Lodato}, G.},
        title = "{Predicting the Kinematic Evidence of Gravitational Instability}",
      journal = {\apj},
         year = 2020,
        month = dec,
       volume = {904},
       number = {2},
          eid = {148},
        pages = {148},
          doi = {10.3847/1538-4357/abac17},
archivePrefix = {arXiv},
       eprint = {2007.15686},
 primaryClass = {astro-ph.SR},
       adsurl = {https://ui.adsabs.harvard.edu/abs/2020ApJ...904..148H}
}

@ARTICLE{Harsono2011,
       author = {{Harsono}, D. and {Alexander}, R.~D. and {Levin}, Y.},
        title = "{Global gravitational instabilities in discs with infall}",
      journal = {\mnras},
         year = 2011,
        month = may,
       volume = {413},
       number = {1},
        pages = {423-433},
          doi = {10.1111/j.1365-2966.2010.18146.x},
archivePrefix = {arXiv},
       eprint = {1012.0724},
 primaryClass = {astro-ph.GA},
       adsurl = {https://ui.adsabs.harvard.edu/abs/2011MNRAS.413..423H}
}

@ARTICLE{Hashimoto2021,
       author = {{Hashimoto}, Jun and {Muto}, Takayuki and {Dong}, Ruobing and {Liu}, Hauyu Baobab and {van der Marel}, Nienke and {Francis}, Logan and {Hasegawa}, Yasuhiro and {Tsukagoshi}, Takashi},
        title = "{ALMA Observations of the Asymmetric Dust Disk around DM Tau}",
      journal = {\apj},
         year = 2021,
        month = apr,
       volume = {911},
       number = {1},
          eid = {5},
        pages = {5},
          doi = {10.3847/1538-4357/abe59f},
archivePrefix = {arXiv},
       eprint = {2102.05905},
 primaryClass = {astro-ph.EP},
       adsurl = {https://ui.adsabs.harvard.edu/abs/2021ApJ...911....5H}
}

@ARTICLE{Huang2018a,
       author = {{Huang}, Jane and {Andrews}, Sean M. and {Dullemond}, Cornelis P. and {Isella}, Andrea and {P{\'e}rez}, Laura M. and {Guzm{\'a}n}, Viviana V. and {{\"O}berg}, Karin I. and {Zhu}, Zhaohuan and {Zhang}, Shangjia and {Bai}, Xue-Ning and {Benisty}, Myriam and {Birnstiel}, Tilman and {Carpenter}, John M. and {Hughes}, A. Meredith and {Ricci}, Luca and {Weaver}, Erik and {Wilner}, David J.},
        title = "{The Disk Substructures at High Angular Resolution Project (DSHARP). II. Characteristics of Annular Substructures}",
      journal = {\apjl},
         year = 2018,
        month = dec,
       volume = {869},
       number = {2},
          eid = {L42},
        pages = {L42},
          doi = {10.3847/2041-8213/aaf740},
archivePrefix = {arXiv},
       eprint = {1812.04041},
 primaryClass = {astro-ph.EP},
       adsurl = {https://ui.adsabs.harvard.edu/abs/2018ApJ...869L..42H}
}

@ARTICLE{Huang2018b,
       author = {{Huang}, Jane and {Andrews}, Sean M. and {P{\'e}rez}, Laura M. and {Zhu}, Zhaohuan and {Dullemond}, Cornelis P. and {Isella}, Andrea and {Benisty}, Myriam and {Bai}, Xue-Ning and {Birnstiel}, Tilman and {Carpenter}, John M. and {Guzm{\'a}n}, Viviana V. and {Hughes}, A. Meredith and {{\"O}berg}, Karin I. and {Ricci}, Luca and {Wilner}, David J. and {Zhang}, Shangjia},
        title = "{The Disk Substructures at High Angular Resolution Project (DSHARP). III. Spiral Structures in the Millimeter Continuum of the Elias 27, IM Lup, and WaOph 6 Disks}",
      journal = {\apjl},
         year = 2018,
        month = dec,
       volume = {869},
       number = {2},
          eid = {L43},
        pages = {L43},
          doi = {10.3847/2041-8213/aaf7a0},
archivePrefix = {arXiv},
       eprint = {1812.04193},
 primaryClass = {astro-ph.SR},
       adsurl = {https://ui.adsabs.harvard.edu/abs/2018ApJ...869L..43H}
}

@ARTICLE{Huang2020,
       author = {{Huang}, Jane and {Andrews}, Sean M. and {{\"O}berg}, Karin I. and {Ansdell}, Megan and {Benisty}, Myriam and {Carpenter}, John M. and {Isella}, Andrea and {P{\'e}rez}, Laura M. and {Ricci}, Luca and {Williams}, Jonathan P. and {Wilner}, David J. and {Zhu}, Zhaohuan},
        title = "{Large-scale CO Spiral Arms and Complex Kinematics Associated with the T Tauri Star RU Lup}",
      journal = {\apj},
         year = 2020,
        month = aug,
       volume = {898},
       number = {2},
          eid = {140},
        pages = {140},
          doi = {10.3847/1538-4357/aba1e1},
archivePrefix = {arXiv},
       eprint = {2007.02974},
 primaryClass = {astro-ph.SR},
       adsurl = {https://ui.adsabs.harvard.edu/abs/2020ApJ...898..140H}
}

@ARTICLE{Huang2021,
       author = {{Huang}, Jane and {Bergin}, Edwin A. and {{\"O}berg}, Karin I. and {Andrews}, Sean M. and {Teague}, Richard and {Law}, Charles J. and {Kalas}, Paul and {Aikawa}, Yuri and {Bae}, Jaehan and {Bergner}, Jennifer B. and {Booth}, Alice S. and {Bosman}, Arthur D. and {Calahan}, Jenny K. and {Cataldi}, Gianni and {Cleeves}, L. Ilsedore and {Czekala}, Ian and {Ilee}, John D. and {Le Gal}, Romane and {Guzm{\'a}n}, Viviana V. and {Long}, Feng and {Loomis}, Ryan A. and {M{\'e}nard}, Fran{\c{c}}ois and {Nomura}, Hideko and {Qi}, Chunhua and {Schwarz}, Kamber R. and {Tsukagoshi}, Takashi and {van't Hoff}, Merel L.~R. and {Walsh}, Catherine and {Wilner}, David J. and {Yamato}, Yoshihide and {Zhang}, Ke},
        title = "{Molecules with ALMA at Planet-forming Scales (MAPS). XIX. Spiral Arms, a Tail, and Diffuse Structures Traced by CO around the GM Aur Disk}",
      journal = {\apjs},
         year = 2021,
        month = nov,
       volume = {257},
       number = {1},
          eid = {19},
        pages = {19},
          doi = {10.3847/1538-4365/ac143e},
archivePrefix = {arXiv},
       eprint = {2109.06224},
 primaryClass = {astro-ph.EP},
       adsurl = {https://ui.adsabs.harvard.edu/abs/2021ApJS..257...19H}
}

@ARTICLE{Huang2025,
       author = {{Huang}, Jane and {Aizawa}, Masataka and {Bae}, Jaehan and {Andrews}, Sean M. and {Benisty}, Myriam and {Bergin}, Edwin A. and {Facchini}, Stefano and {Ginski}, Christian and {K{\"u}ffmeier}, Michael},
        title = "{Grand Design Spiral Arms in the Compact, Embedded Protoplanetary Disk of Haro 6-13}",
      journal = {\apj},
         year = 2025,
        month = jul,
       volume = {988},
       number = {1},
          eid = {106},
        pages = {106},
          doi = {10.3847/1538-4357/ade1d8},
archivePrefix = {arXiv},
       eprint = {2506.06433},
 primaryClass = {astro-ph.EP},
       adsurl = {https://ui.adsabs.harvard.edu/abs/2025ApJ...988..106H}
}

@ARTICLE{Huhn2026,
       author = {{H{\"u}hn}, L.-A. and {Kimmig}, C.~N. and {Dullemond}, C.~P.},
        title = "{Spiral formation caused by late infall onto protoplanetary disks}",
      journal = {\aap},
         year = 2026,
        month = apr,
       volume = {708},
          eid = {A93},
        pages = {A93},
          doi = {10.1051/0004-6361/202558773},
archivePrefix = {arXiv},
       eprint = {2603.03442},
 primaryClass = {astro-ph.EP},
       adsurl = {https://ui.adsabs.harvard.edu/abs/2026A&A...708A..93H}
}

@ARTICLE{Hunter2007,
       author = {{Hunter}, John D.},
        title = "{Matplotlib: A 2D Graphics Environment}",
      journal = {Computing in Science and Engineering},
         year = 2007,
        month = may,
       volume = {9},
       number = {3},
        pages = {90-95},
          doi = {10.1109/MCSE.2007.55},
       adsurl = {https://ui.adsabs.harvard.edu/abs/2007CSE.....9...90H}
}

@ARTICLE{Jennings2020,
       author = {{Jennings}, Jeff and {Booth}, Richard A. and {Tazzari}, Marco and {Rosotti}, Giovanni P. and {Clarke}, Cathie J.},
        title = "{frankenstein: protoplanetary disc brightness profile reconstruction at sub-beam resolution with a rapid Gaussian process}",
      journal = {\mnras},
         year = 2020,
        month = jul,
       volume = {495},
       number = {3},
        pages = {3209-3232},
          doi = {10.1093/mnras/staa1365},
archivePrefix = {arXiv},
       eprint = {2005.07709},
 primaryClass = {astro-ph.EP},
       adsurl = {https://ui.adsabs.harvard.edu/abs/2020MNRAS.495.3209J}
}

@ARTICLE{Izquierdo2021,
       author = {{Izquierdo}, A.~F. and {Testi}, L. and {Facchini}, S. and {Rosotti}, G.~P. and {van Dishoeck}, E.~F.},
        title = "{The Disc Miner. I. A statistical framework to detect and quantify kinematical perturbations driven by young planets in discs}",
      journal = {\aap},
         year = 2021,
        month = jun,
       volume = {650},
          eid = {A179},
        pages = {A179},
          doi = {10.1051/0004-6361/202140779},
archivePrefix = {arXiv},
       eprint = {2104.09596},
 primaryClass = {astro-ph.EP},
       adsurl = {https://ui.adsabs.harvard.edu/abs/2021A&A...650A.179I}
}

@ARTICLE{Jennings2022a,
       author = {{Jennings}, Jeff and {Tazzari}, Marco and {Clarke}, Cathie J. and {Booth}, Richard A. and {Rosotti}, Giovanni P.},
        title = "{Superresolution trends in the ALMA Taurus survey: structured inner discs and compact discs}",
      journal = {\mnras},
         year = 2022,
        month = aug,
       volume = {514},
       number = {4},
        pages = {6053-6073},
          doi = {10.1093/mnras/stac1770},
archivePrefix = {arXiv},
       eprint = {2206.11308},
 primaryClass = {astro-ph.EP},
       adsurl = {https://ui.adsabs.harvard.edu/abs/2022MNRAS.514.6053J}
}

@ARTICLE{Jennings2022b,
       author = {{Jennings}, Jeff and {Booth}, Richard A. and {Tazzari}, Marco and {Clarke}, Cathie J. and {Rosotti}, Giovanni P.},
        title = "{A super-resolution analysis of the DSHARP survey: substructure is common in the inner 30 au}",
      journal = {\mnras},
         year = 2022,
        month = jan,
       volume = {509},
       number = {2},
        pages = {2780-2799},
          doi = {10.1093/mnras/stab3185},
archivePrefix = {arXiv},
       eprint = {2103.02392},
 primaryClass = {astro-ph.EP},
       adsurl = {https://ui.adsabs.harvard.edu/abs/2022MNRAS.509.2780J}
}

@INCOLLECTION{Kluyver2016,
       author = {{Kluyver}, Thomas and {Ragan-Kelley}, Benjain and {P{\'e}rez}, Fernando and {Granger}, Brian and {Bussonnier}, Matthias and {Frederic}, Jonathan and {Kelley}, Kyle and {Hamrick}, Jessica and {Grout}, Jason and {Corlay}, Sylvain and {Ivanov}, Paul and {Avila}, Dami{\'a}n and {Abdalla}, Safia and {Willing}, Carol and {Jupyter Development Team}},
        title = "{Jupyter Notebooks{\textemdash}a publishing format for reproducible computational workflows}",
    booktitle = {IOS Press},
         year = 2016,
        pages = {87-90},
          doi = {10.3233/978-1-61499-649-1-87},
       adsurl = {https://ui.adsabs.harvard.edu/abs/2016ppap.book...87K}
}

@ARTICLE{Kratter2016,
       author = {{Kratter}, Kaitlin and {Lodato}, Giuseppe},
        title = "{Gravitational Instabilities in Circumstellar Disks}",
      journal = {\araa},
         year = 2016,
        month = sep,
       volume = {54},
        pages = {271-311},
          doi = {10.1146/annurev-astro-081915-023307},
archivePrefix = {arXiv},
       eprint = {1603.01280},
 primaryClass = {astro-ph.SR},
       adsurl = {https://ui.adsabs.harvard.edu/abs/2016ARA&A..54..271K}
}

@ARTICLE{Lee2020,
       author = {{Lee}, Chin-Fei and {Li}, Zhi-Yun and {Turner}, Neal J.},
        title = "{Spiral structures in an embedded protostellar disk driven by envelope accretion}",
      journal = {Nature Astronomy},
         year = 2020,
        month = feb,
       volume = {4},
        pages = {142-146},
          doi = {10.1038/s41550-019-0905-x},
archivePrefix = {arXiv},
       eprint = {1909.02812},
 primaryClass = {astro-ph.GA},
       adsurl = {https://ui.adsabs.harvard.edu/abs/2020NatAs...4..142L}
}

@ARTICLE{Lodato2023,
       author = {{Lodato}, G. and {Rampinelli}, L. and {Viscardi}, E. and {Longarini}, C. and {Izquierdo}, A. and {Paneque-Carre{\~n}o}, T. and {Testi}, L. and {Facchini}, S. and {Miotello}, A. and {Veronesi}, B. and {Hall}, C.},
        title = "{Dynamical mass measurements of two protoplanetary discs}",
      journal = {\mnras},
         year = 2023,
        month = jan,
       volume = {518},
       number = {3},
        pages = {4481-4493},
          doi = {10.1093/mnras/stac3223},
archivePrefix = {arXiv},
       eprint = {2211.03712},
 primaryClass = {astro-ph.EP},
       adsurl = {https://ui.adsabs.harvard.edu/abs/2023MNRAS.518.4481L}
}

@ARTICLE{Longarini2025,
       author = {{Longarini}, Cristiano and {Price}, Daniel J. and {Kratter}, Kaitlin M. and {Lodato}, Giuseppe and {Clarke}, Cathie J.},
        title = "{Infall-driven gravitational instability in accretion discs}",
      journal = {\mnras},
         year = 2025,
        month = aug,
       volume = {541},
       number = {2},
        pages = {1145-1163},
          doi = {10.1093/mnras/staf1018},
archivePrefix = {arXiv},
       eprint = {2506.13701},
 primaryClass = {astro-ph.EP},
       adsurl = {https://ui.adsabs.harvard.edu/abs/2025MNRAS.541.1145L}
}

@ARTICLE{Martire2024,
       author = {{Martire}, P. and {Longarini}, C. and {Lodato}, G. and {Rosotti}, G.~P. and {Winter}, A. and {Facchini}, S. and {Hardiman}, C. and {Benisty}, M. and {Stadler}, J. and {Izquierdo}, A.~F. and {Testi}, Leonardo},
        title = "{Rotation curves in protoplanetary disks with thermal stratification. Physical model and observational evidence in MAPS disks}",
      journal = {\aap},
         year = 2024,
        month = jun,
       volume = {686},
          eid = {A9},
        pages = {A9},
          doi = {10.1051/0004-6361/202348546},
archivePrefix = {arXiv},
       eprint = {2402.12236},
 primaryClass = {astro-ph.EP},
       adsurl = {https://ui.adsabs.harvard.edu/abs/2024A&A...686A...9M}
}

@ARTICLE{Longarini2026,
       author = {{Longarini}, Cristiano and {Lodato}, Giuseppe},
        title = "{Gravitational instability in planet-forming discs}",
      journal = {arXiv e-prints},
         year = 2026,
        month = sep,
          eid = {arXiv:2609.05080},
        pages = {arXiv:2609.05080},
          doi = {10.48550/arXiv.2609.05080},
archivePrefix = {arXiv},
       eprint = {2609.05080},
 primaryClass = {astro-ph.EP},
       adsurl = {https://ui.adsabs.harvard.edu/abs/2026arXiv260905080L}
}

@InProceedings{Mckinney-proc-scipy-2010,
  author    = { Wes McKinney },
  title     = { Data Structures for Statistical Computing in Python },
  booktitle = { Proceedings of the 9th Python in Science Conference },
  pages     = { 51 - 56 },
  year      = { 2010 },
  editor    = { St\'efan van der Walt and Jarrod Millman }
}

@INPROCEEDINGS{McMullin2007,
       author = {{McMullin}, J.~P. and {Waters}, B. and {Schiebel}, D. and {Young}, W. and {Golap}, K.},
        title = "{CASA Architecture and Applications}",
    booktitle = {Astronomical Data Analysis Software and Systems XVI},
         year = 2007,
       editor = {{Shaw}, R.~A. and {Hill}, F. and {Bell}, D.~J.},
       series = {Astronomical Society of the Pacific Conference Series},
       volume = {376},
        month = oct,
        pages = {127},
       adsurl = {https://ui.adsabs.harvard.edu/abs/2007ASPC..376..127M}
}

@ARTICLE{Montesinos2016,
       author = {{Montesinos}, Mat{\'\i}as and {Perez}, Sebastian and {Casassus}, Simon and {Marino}, Sebastian and {Cuadra}, Jorge and {Christiaens}, Valentin},
        title = "{Spiral Waves Triggered by Shadows in Transition Disks}",
      journal = {\apjl},
         year = 2016,
        month = may,
       volume = {823},
       number = {1},
          eid = {L8},
        pages = {L8},
          doi = {10.3847/2041-8205/823/1/L8},
archivePrefix = {arXiv},
       eprint = {1601.07912},
 primaryClass = {astro-ph.EP},
       adsurl = {https://ui.adsabs.harvard.edu/abs/2016ApJ...823L...8M}
}

@ARTICLE{Muto2012,
       author = {{Muto}, T. and {Grady}, C.~A. and {Hashimoto}, J. and {Fukagawa}, M. and {Hornbeck}, J.~B. and {Sitko}, M. and {Russell}, R. and {Werren}, C. and {Cur{\'e}}, M. and {Currie}, T. and {Ohashi}, N. and {Okamoto}, Y. and {Momose}, M. and {Honda}, M. and {Inutsuka}, S. and {Takeuchi}, T. and {Dong}, R. and {Abe}, L. and {Brandner}, W. and {Brandt}, T. and {Carson}, J. and {Egner}, S. and {Feldt}, M. and {Fukue}, T. and {Goto}, M. and {Guyon}, O. and {Hayano}, Y. and {Hayashi}, M. and {Hayashi}, S. and {Henning}, T. and {Hodapp}, K.~W. and {Ishii}, M. and {Iye}, M. and {Janson}, M. and {Kandori}, R. and {Knapp}, G.~R. and {Kudo}, T. and {Kusakabe}, N. and {Kuzuhara}, M. and {Matsuo}, T. and {Mayama}, S. and {McElwain}, M.~W. and {Miyama}, S. and {Morino}, J.-I. and {Moro-Martin}, A. and {Nishimura}, T. and {Pyo}, T.-S. and {Serabyn}, E. and {Suto}, H. and {Suzuki}, R. and {Takami}, M. and {Takato}, N. and {Terada}, H. and {Thalmann}, C. and {Tomono}, D. and {Turner}, E.~L. and {Watanabe}, M. and {Wisniewski}, J.~P. and {Yamada}, T. and {Takami}, H. and {Usuda}, T. and {Tamura}, M.},
        title = "{Discovery of Small-scale Spiral Structures in the Disk of SAO 206462 (HD 135344B): Implications for the Physical State of the Disk from Spiral Density Wave Theory}",
      journal = {\apjl},
         year = 2012,
        month = apr,
       volume = {748},
       number = {2},
          eid = {L22},
        pages = {L22},
          doi = {10.1088/2041-8205/748/2/L22},
archivePrefix = {arXiv},
       eprint = {1202.6139},
 primaryClass = {astro-ph.EP},
       adsurl = {https://ui.adsabs.harvard.edu/abs/2012ApJ...748L..22M}
}

@ARTICLE{Oberg2021,
       author = {{{\"O}berg}, Karin I. and {Guzm{\'a}n}, Viviana V. and {Walsh}, Catherine and {Aikawa}, Yuri and {Bergin}, Edwin A. and {Law}, Charles J. and {Loomis}, Ryan A. and {Alarc{\'o}n}, Felipe and {Andrews}, Sean M. and {Bae}, Jaehan and {Bergner}, Jennifer B. and {Boehler}, Yann and {Booth}, Alice S. and {Bosman}, Arthur D. and {Calahan}, Jenny K. and {Cataldi}, Gianni and {Cleeves}, L. Ilsedore and {Czekala}, Ian and {Furuya}, Kenji and {Huang}, Jane and {Ilee}, John D. and {Kurtovic}, Nicolas T. and {Le Gal}, Romane and {Liu}, Yao and {Long}, Feng and {M{\'e}nard}, Fran{\c{c}}ois and {Nomura}, Hideko and {P{\'e}rez}, Laura M. and {Qi}, Chunhua and {Schwarz}, Kamber R. and {Sierra}, Anibal and {Teague}, Richard and {Tsukagoshi}, Takashi and {Yamato}, Yoshihide and {van't Hoff}, Merel L.~R. and {Waggoner}, Abygail R. and {Wilner}, David J. and {Zhang}, Ke},
        title = "{Molecules with ALMA at Planet-forming Scales (MAPS). I. Program Overview and Highlights}",
      journal = {\apjs},
         year = 2021,
        month = nov,
       volume = {257},
       number = {1},
          eid = {1},
        pages = {1},
          doi = {10.3847/1538-4365/ac1432},
archivePrefix = {arXiv},
       eprint = {2109.06268},
 primaryClass = {astro-ph.EP},
       adsurl = {https://ui.adsabs.harvard.edu/abs/2021ApJS..257....1O}
}

@ARTICLE{Ogilvie2002,
       author = {{Ogilvie}, G.~I. and {Lubow}, S.~H.},
        title = "{On the wake generated by a planet in a disc}",
      journal = {\mnras},
         year = 2002,
        month = mar,
       volume = {330},
       number = {4},
        pages = {950-954},
          doi = {10.1046/j.1365-8711.2002.05148.x},
archivePrefix = {arXiv},
       eprint = {astro-ph/0111265},
 primaryClass = {astro-ph},
       adsurl = {https://ui.adsabs.harvard.edu/abs/2002MNRAS.330..950O}
}

@ARTICLE{Paneque2024,
       author = {{Paneque-Carre{\~n}o}, T. and {Izquierdo}, A.~F. and {Teague}, R. and {Miotello}, A. and {Bergin}, E.~A. and {Loomis}, R. and {van Dishoeck}, E.~F.},
        title = "{High turbulence in the IM Lup protoplanetary disk. Direct observational constraints from CN and C$_{2}$H emission}",
      journal = {\aap},
         year = 2024,
        month = apr,
       volume = {684},
          eid = {A174},
        pages = {A174},
          doi = {10.1051/0004-6361/202347757},
archivePrefix = {arXiv},
       eprint = {2312.04618},
 primaryClass = {astro-ph.EP},
       adsurl = {https://ui.adsabs.harvard.edu/abs/2024A&A...684A.174P}
}

@ARTICLE{Perez2015,
       author = {{Perez}, Sebastian and {Dunhill}, A. and {Casassus}, S. and {Roman}, P. and {Szul{\'a}gyi}, J. and {Flores}, C. and {Marino}, S. and {Montesinos}, M.},
        title = "{Planet Formation Signposts: Observability of Circumplanetary Disks via Gas Kinematics}",
      journal = {\apjl},
         year = 2015,
        month = sep,
       volume = {811},
       number = {1},
          eid = {L5},
        pages = {L5},
          doi = {10.1088/2041-8205/811/1/L5},
archivePrefix = {arXiv},
       eprint = {1505.06808},
 primaryClass = {astro-ph.EP},
       adsurl = {https://ui.adsabs.harvard.edu/abs/2015ApJ...811L...5P}
}

@ARTICLE{perez2016,
       author = {{P{\'e}rez}, Laura M. and {Carpenter}, John M. and {Andrews}, Sean M. and {Ricci}, Luca and {Isella}, Andrea and {Linz}, Hendrik and {Sargent}, Anneila I. and {Wilner}, David J. and {Henning}, Thomas and {Deller}, Adam T. and {Chandler}, Claire J. and {Dullemond}, Cornelis P. and {Lazio}, Joseph and {Menten}, Karl M. and {Corder}, Stuartt A. and {Storm}, Shaye and {Testi}, Leonardo and {Tazzari}, Marco and {Kwon}, Woojin and {Calvet}, Nuria and {Greaves}, Jane S. and {Harris}, Robert J. and {Mundy}, Lee G.},
        title = "{Spiral density waves in a young protoplanetary disk}",
      journal = {Science},
         year = 2016,
        month = sep,
       volume = {353},
       number = {6307},
        pages = {1519-1521},
          doi = {10.1126/science.aaf8296},
archivePrefix = {arXiv},
       eprint = {1610.05139},
 primaryClass = {astro-ph.GA},
       adsurl = {https://ui.adsabs.harvard.edu/abs/2016Sci...353.1519P}
}

@ARTICLE{Pezzotta2026,
       author = {{Pezzotta}, V. and {Facchini}, S. and {Izquierdo}, A.~F. and {Lodato}, G. and {Longarini}, C. and {Bae}, J. and {Galloway-Sprietsma}, M. and {Pinte}, C. and {Law}, C.~J. and {Paneque-Carre{\~n}o}, T.},
        title = "{Extending dynamical mass measurements: probing GI as a possible origin of mm-dust spirals}",
      journal = {arXiv e-prints},
         year = 2026,
        month = jul,
          eid = {arXiv:2607.15923},
        pages = {arXiv:2607.15923},
          doi = {10.48550/arXiv.2607.15923},
archivePrefix = {arXiv},
       eprint = {2607.15923},
 primaryClass = {astro-ph.EP},
       adsurl = {https://ui.adsabs.harvard.edu/abs/2026arXiv260715923P}
}

@ARTICLE{Pinte2018,
       author = {{Pinte}, C. and {M{\'e}nard}, F. and {Duch{\^e}ne}, G. and {Hill}, T. and {Dent}, W.~R.~F. and {Woitke}, P. and {Maret}, S. and {van der Plas}, G. and {Hales}, A. and {Kamp}, I. and {Thi}, W.~F. and {de Gregorio-Monsalvo}, I. and {Rab}, C. and {Quanz}, S.~P. and {Avenhaus}, H. and {Carmona}, A. and {Casassus}, S.},
        title = "{Direct mapping of the temperature and velocity gradients in discs. Imaging the vertical CO snow line around IM Lupi}",
      journal = {\aap},
         year = 2018,
        month = jan,
       volume = {609},
          eid = {A47},
        pages = {A47},
          doi = {10.1051/0004-6361/201731377},
archivePrefix = {arXiv},
       eprint = {1710.06450},
 primaryClass = {astro-ph.SR},
       adsurl = {https://ui.adsabs.harvard.edu/abs/2018A&A...609A..47P}
}

@ARTICLE{Pinte2020,
       author = {{Pinte}, C. and {Price}, D.~J. and {M{\'e}nard}, F. and {Duch{\^e}ne}, G. and {Christiaens}, V. and {Andrews}, S.~M. and {Huang}, J. and {Hill}, T. and {van der Plas}, G. and {Perez}, L.~M. and {Isella}, A. and {Boehler}, Y. and {Dent}, W.~R.~F. and {Mentiplay}, D. and {Loomis}, R.~A.},
        title = "{Nine Localized Deviations from Keplerian Rotation in the DSHARP Circumstellar Disks: Kinematic Evidence for Protoplanets Carving the Gaps}",
      journal = {\apjl},
         year = 2020,
        month = feb,
       volume = {890},
       number = {1},
          eid = {L9},
        pages = {L9},
          doi = {10.3847/2041-8213/ab6dda},
archivePrefix = {arXiv},
       eprint = {2001.07720},
 primaryClass = {astro-ph.SR},
       adsurl = {https://ui.adsabs.harvard.edu/abs/2020ApJ...890L...9P}
}

@ARTICLE{Ren2020,
       author = {{Ren}, Bin and {Dong}, Ruobing and {van Holstein}, Rob G. and {Ruffio}, Jean-Baptiste and {Calvin}, Benjamin A. and {Girard}, Julien H. and {Benisty}, Myriam and {Boccaletti}, Anthony and {Esposito}, Thomas M. and {Choquet}, {\'E}lodie and {Mawet}, Dimitri and {Pueyo}, Laurent and {Stolker}, Tomas and {Chiang}, Eugenede and {Boer}, Jozua and {Debes}, John H. and {Garufi}, Antonio and {Grady}, Carol A. and {Hines}, Dean C. and {Maire}, Anne-Lise and {M{\'e}nard}, Fran{\c{c}}ois and {Millar-Blanchaer}, Maxwell A. and {Perrin}, Marshall D. and {Poteet}, Charles A. and {Schneider}, Glenn},
        title = "{Dynamical Evidence of a Spiral Arm-driving Planet in the MWC 758 Protoplanetary Disk}",
      journal = {\apjl},
         year = 2020,
        month = aug,
       volume = {898},
       number = {2},
          eid = {L38},
        pages = {L38},
          doi = {10.3847/2041-8213/aba43e},
archivePrefix = {arXiv},
       eprint = {2007.04980},
 primaryClass = {astro-ph.EP},
       adsurl = {https://ui.adsabs.harvard.edu/abs/2020ApJ...898L..38R}
}

@ARTICLE{Ren2024,
       author = {{Ren}, Bin B. and {Xie}, Chen and {Benisty}, Myriam and {Dong}, Ruobing and {Bae}, Jaehan and {Stolker}, Tomas and {van Holstein}, Rob G. and {Debes}, John H. and {Garufi}, Antonio and {Ginski}, Christian and {Kraus}, Stefan},
        title = "{A companion in V1247 Ori supported by motion in the pattern of the spiral arm}",
      journal = {\aap},
         year = 2024,
        month = jan,
       volume = {681},
          eid = {L2},
        pages = {L2},
          doi = {10.1051/0004-6361/202348114},
archivePrefix = {arXiv},
       eprint = {2310.15430},
 primaryClass = {astro-ph.EP},
       adsurl = {https://ui.adsabs.harvard.edu/abs/2024A&A...681L...2R}
}

@ARTICLE{Rowther2025,
       author = {{Rowther}, Sahl and {Nealon}, Rebecca and {Alexander}, Richard and {Meru}, Farzana},
        title = "{Leading and trailing spiral arms in a nearly broken protoplanetary disc}",
      journal = {\mnras},
         year = 2025,
        month = sep,
       volume = {542},
       number = {2},
        pages = {1430-1442},
          doi = {10.1093/mnras/staf1301},
archivePrefix = {arXiv},
       eprint = {2508.04914},
 primaryClass = {astro-ph.EP},
       adsurl = {https://ui.adsabs.harvard.edu/abs/2025MNRAS.542.1430R}
}

@ARTICLE{Speedie2022,
       author = {{Speedie}, Jessica and {Dong}, Ruobing},
        title = "{Testing Velocity Kinks as a Planet Detection Method: Do Velocity Kinks in Surface Gas Emission Trace Planetary Spiral Wakes in the Midplane Continuum?}",
      journal = {\apjl},
         year = 2022,
        month = dec,
       volume = {940},
       number = {2},
          eid = {L43},
        pages = {L43},
          doi = {10.3847/2041-8213/aca074},
archivePrefix = {arXiv},
       eprint = {2211.05757},
 primaryClass = {astro-ph.EP},
       adsurl = {https://ui.adsabs.harvard.edu/abs/2022ApJ...940L..43S}
}

@ARTICLE{Speedie2025,
       author = {{Speedie}, Jessica and {Dong}, Ruobing and {Teague}, Richard and {Segura-Cox}, Dominique and {Pineda}, Jaime E. and {Calcino}, Josh and {Longarini}, Cristiano and {Hall}, Cassandra and {Tang}, Ya-Wen and {Hashimoto}, Jun and {Paneque-Carre{\~n}o}, Teresa and {Lodato}, Giuseppe and {Veronesi}, Bennedetta},
        title = "{Mapping the Merging Zone of Late Infall in the AB Aur Planet-forming System}",
      journal = {\apjl},
         year = 2025,
        month = mar,
       volume = {981},
       number = {2},
          eid = {L30},
        pages = {L30},
          doi = {10.3847/2041-8213/adb7d5},
archivePrefix = {arXiv},
       eprint = {2503.01957},
 primaryClass = {astro-ph.EP},
       adsurl = {https://ui.adsabs.harvard.edu/abs/2025ApJ...981L..30S}
}

@ARTICLE{Su2024,
       author = {{Su}, Zehao and {Bai}, Xue-Ning},
        title = "{Dynamical Consequence of Shadows Cast to the Outer Protoplanetary Disks. I. Two-dimensional Simulations}",
      journal = {\apj},
         year = 2024,
        month = nov,
       volume = {975},
       number = {1},
          eid = {126},
        pages = {126},
          doi = {10.3847/1538-4357/ad7581},
archivePrefix = {arXiv},
       eprint = {2407.12659},
 primaryClass = {astro-ph.EP},
       adsurl = {https://ui.adsabs.harvard.edu/abs/2024ApJ...975..126S}
}

@ARTICLE{Su2026,
       author = {{Su}, Wei-Shan and {Smallwood}, Jeremy L. and {Lin}, Min-Kai and {Yang}, Chao-Chin and {Cuello}, Nicol{\'a}s},
        title = "{Evolution of dust in a protoplanetary disc driven by stellar flybys: implications for the streaming instability}",
      journal = {\mnras},
         year = 2026,
        month = apr,
       volume = {547},
       number = {2},
          eid = {stag173},
        pages = {stag173},
          doi = {10.1093/mnras/stag173},
archivePrefix = {arXiv},
       eprint = {2601.17278},
 primaryClass = {astro-ph.EP},
       adsurl = {https://ui.adsabs.harvard.edu/abs/2026MNRAS.547ag173S}
}

@ARTICLE{Teague2021,
       author = {{Teague}, Richard and {Bae}, Jaehan and {Aikawa}, Yuri and {Andrews}, Sean M. and {Bergin}, Edwin A. and {Bergner}, Jennifer B. and {Boehler}, Yann and {Booth}, Alice S. and {Bosman}, Arthur D. and {Cataldi}, Gianni and {Czekala}, Ian and {Guzm{\'a}n}, Viviana V. and {Huang}, Jane and {Ilee}, John D. and {Law}, Charles J. and {Le Gal}, Romane and {Long}, Feng and {Loomis}, Ryan A. and {M{\'e}nard}, Fran{\c{c}}ois and {{\"O}berg}, Karin I. and {P{\'e}rez}, Laura M. and {Schwarz}, Kamber R. and {Sierra}, Anibal and {Walsh}, Catherine and {Wilner}, David J. and {Yamato}, Yoshihide and {Zhang}, Ke},
        title = "{Molecules with ALMA at Planet-forming Scales (MAPS). XVIII. Kinematic Substructures in the Disks of HD 163296 and MWC 480}",
      journal = {\apjs},
         year = 2021,
        month = nov,
       volume = {257},
       number = {1},
          eid = {18},
        pages = {18},
          doi = {10.3847/1538-4365/ac1438},
archivePrefix = {arXiv},
       eprint = {2109.06218},
 primaryClass = {astro-ph.EP},
       adsurl = {https://ui.adsabs.harvard.edu/abs/2021ApJS..257...18T}
}

@ARTICLE{Tomida2017,
       author = {{Tomida}, K. and {Machida}, M.~N. and {Hosokawa}, T. and {Sakurai}, Y. and {Lin}, C.~H.},
        title = "{Grand design spiral arms in a young forming circumstellar disk .}",
      journal = {\memsai},
         year = 2017,
        month = jan,
       volume = {88},
        pages = {822},
       adsurl = {https://ui.adsabs.harvard.edu/abs/2017MmSAI..88..822T}
}

@ARTICLE{Ueda2024,
       author = {{Ueda}, Takahiro and {Tazaki}, Ryo and {Okuzumi}, Satoshi and {Flock}, Mario and {Sudarshan}, Prakruti},
        title = "{Support for fragile porous dust in a gravitationally self-regulated disk around IM Lup}",
      journal = {Nature Astronomy},
         year = 2024,
        month = sep,
       volume = {8},
        pages = {1148-1158},
          doi = {10.1038/s41550-024-02308-6},
archivePrefix = {arXiv},
       eprint = {2406.07427},
 primaryClass = {astro-ph.EP},
       adsurl = {https://ui.adsabs.harvard.edu/abs/2024NatAs...8.1148U}
}

@ARTICLE{Verrios2022,
       author = {{Verrios}, Harrison J. and {Price}, Daniel J. and {Pinte}, Christophe and {Hilder}, Thomas and {Calcino}, Josh},
        title = "{Kinematic Evidence for an Embedded Planet in the IM Lupi Disk}",
      journal = {\apjl},
         year = 2022,
        month = jul,
       volume = {934},
       number = {1},
          eid = {L11},
        pages = {L11},
          doi = {10.3847/2041-8213/ac7f44},
archivePrefix = {arXiv},
       eprint = {2207.02869},
 primaryClass = {astro-ph.EP},
       adsurl = {https://ui.adsabs.harvard.edu/abs/2022ApJ...934L..11V}
}

@ARTICLE{Veronesi2021,
       author = {{Veronesi}, Benedetta and {Paneque-Carre{\~n}o}, Teresa and {Lodato}, Giuseppe and {Testi}, Leonardo and {P{\'e}rez}, Laura M. and {Bertin}, Giuseppe and {Hall}, Cassandra},
        title = "{A Dynamical Measurement of the Disk Mass in Elias 227}",
      journal = {\apjl},
         year = 2021,
        month = jun,
       volume = {914},
       number = {2},
          eid = {L27},
        pages = {L27},
          doi = {10.3847/2041-8213/abfe6a},
archivePrefix = {arXiv},
       eprint = {2104.09530},
 primaryClass = {astro-ph.EP},
       adsurl = {https://ui.adsabs.harvard.edu/abs/2021ApJ...914L..27V}
}

@ARTICLE{Virtanen2020,
       author = {{Virtanen}, Pauli and {Gommers}, Ralf and {Oliphant}, Travis E. and {Haberland}, Matt and {Reddy}, Tyler and {Cournapeau}, David and {Burovski}, Evgeni and {Peterson}, Pearu and {Weckesser}, Warren and {Bright}, Jonathan and {van der Walt}, St{\'e}fan J. and {Brett}, Matthew and {Wilson}, Joshua and {Millman}, K. Jarrod and {Mayorov}, Nikolay and {Nelson}, Andrew R.~J. and {Jones}, Eric and {Kern}, Robert and {Larson}, Eric and {Carey}, C.~J. and {Polat}, {\.I}lhan and {Feng}, Yu and {Moore}, Eric W. and {VanderPlas}, Jake and {Laxalde}, Denis and {Perktold}, Josef and {Cimrman}, Robert and {Henriksen}, Ian and {Quintero}, E.~A. and {Harris}, Charles R. and {Archibald}, Anne M. and {Ribeiro}, Ant{\^o}nio H. and {Pedregosa}, Fabian and {van Mulbregt}, Paul and {SciPy 1. 0 Contributors}},
        title = "{SciPy 1.0: fundamental algorithms for scientific computing in Python}",
      journal = {Nature Methods},
         year = 2020,
        month = feb,
       volume = {17},
        pages = {261-272},
          doi = {10.1038/s41592-019-0686-2},
archivePrefix = {arXiv},
       eprint = {1907.10121},
 primaryClass = {cs.MS},
       adsurl = {https://ui.adsabs.harvard.edu/abs/2020NatMe..17..261V}
}

@ARTICLE{Walt2011,
       author = {{van der Walt}, St{\'e}fan and {Colbert}, S. Chris and {Varoquaux}, Ga{\"e}l},
        title = "{The NumPy Array: A Structure for Efficient Numerical Computation}",
      journal = {Computing in Science and Engineering},
         year = 2011,
        month = mar,
       volume = {13},
       number = {2},
        pages = {22-30},
          doi = {10.1109/MCSE.2011.37},
archivePrefix = {arXiv},
       eprint = {1102.1523},
 primaryClass = {cs.MS},
       adsurl = {https://ui.adsabs.harvard.edu/abs/2011CSE....13b..22V}
}

@ARTICLE{Winter2024,
       author = {{Winter}, Andrew J. and {Benisty}, Myriam and {Manara}, Carlo F. and {Gupta}, Aashish},
        title = "{Spatially correlated stellar accretion in the Lupus star-forming region: Evidence for ongoing infall from the interstellar medium}",
      journal = {\aap},
         year = 2024,
        month = nov,
       volume = {691},
          eid = {A169},
        pages = {A169},
          doi = {10.1051/0004-6361/202452120},
archivePrefix = {arXiv},
       eprint = {2409.17220},
 primaryClass = {astro-ph.EP},
       adsurl = {https://ui.adsabs.harvard.edu/abs/2024A&A...691A.169W}
}

@ARTICLE{Xie2024,
       author = {{Xie}, Chen and {Xie}, Chengyan and {Ren}, Bin B. and {Benisty}, Myriam and {Ginski}, Christian and {Fang}, Taotao and {Casassus}, Simon and {Bae}, Jaehan and {Facchini}, Stefano and {M{\'e}nard}, Fran{\c{c}}ois and {van Holstein}, Rob G.},
        title = "{Disk Evolution Study Through Imaging of Nearby Young Stars (DESTINYS): Dynamical Evidence of a Spiral-Arm-Driving and Gap-Opening Protoplanet from SAO 206462 Spiral Motion}",
      journal = {Universe},
         year = 2024,
        month = dec,
       volume = {10},
       number = {12},
          eid = {465},
        pages = {465},
          doi = {10.3390/universe10120465},
archivePrefix = {arXiv},
       eprint = {2412.14402},
 primaryClass = {astro-ph.EP},
       adsurl = {https://ui.adsabs.harvard.edu/abs/2024Univ...10..465X}
}

@ARTICLE{Yoshida2025,
       author = {{Yoshida}, Tomohiro C. and {Nomura}, Hideko and {Doi}, Kiyoaki and {Barraza-Alfaro}, Marcelo and {Teague}, Richard and {Furuya}, Kenji and {Yamato}, Yoshihide and {Tsukagoshi}, Takashi},
        title = "{Winding motion of spirals in a gravitationally unstable protoplanetary disk}",
      journal = {Nature Astronomy},
         year = 2025,
        month = nov,
       volume = {9},
       number = {11},
        pages = {1672-1679},
          doi = {10.1038/s41550-025-02639-y},
archivePrefix = {arXiv},
       eprint = {2509.19761},
 primaryClass = {astro-ph.EP},
       adsurl = {https://ui.adsabs.harvard.edu/abs/2025NatAs...9.1672Y}
}

\appendix
\section{Uncertainty Estimation for the Pattern Speed}
\label{app:Uncertainty}
In an interferometric image, the intensity in each pixel dose not represent an independent measurement, because the image is smoothed by the synthesized beam. As a result, the noise in neighboring pixels is spatially correlated on approximately the beam scale and not independent.

To account for this effect, we model the observed residual-image noise as a
beam-smoothed realization of an underlying uncorrelated noise field.
Along a ring at radius $r$, we approximate the action of the circular
synthesized beam as a one-dimensional Gaussian convolution in the azimuthal
direction. We then propagate this correlated noise into the uncertainty of
the Fourier phase and, subsequently, the pattern speed.

All residual images used in the pattern speed analysis are convolved with a
common circular Gaussian beam with $B_{\rm FWHM}=150$~mas.
At radius $r$, the beam corresponds approximately to an angular scale
$B_{\phi}(r)\simeq B_{\rm FWHM}/r$ along the ring.
The standard deviation of the corresponding Gaussian smoothing kernel is
therefore
$\sigma_{\rm b}(r)=B_{\phi}(r)/(2\sqrt{2\ln2})$.
For $N_\phi$ uniformly spaced azimuthal samples, with
$\Delta\theta=2\pi/N_\phi$, we define the normalized Gaussian smoothing
kernel as
\begin{equation}
g_j(r)
\propto
\exp\left(-\frac{\Delta\theta_j^2}{2\sigma_{\rm b}^2(r)}\right),
\qquad
\sum_j g_j(r)=1,
\label{eq:noise_kernel}
\end{equation}
where $j$ indexes the azimuthal samples of the kernel and
$\Delta\theta_j$ is the minimum angular separation from the kernel center
on the periodic ring.

Let $\sigma_0(r)$ denote the standard deviation of the underlying
uncorrelated noise field.
Convolution with the normalized kernel gives a variance
$\sigma_0^2(r)\sum_j g_j^2(r)$.
Equating the corresponding rms to the measured off-source image rms,
$\sigma_I$, gives
\begin{equation}
\sigma_0(r)
=
\frac{\sigma_I}
{\sqrt{\sum_j g_j^2(r)}}.
\label{eq:white_noise_rms}
\end{equation}

For the underlying uncorrelated noise sampled at $N_\phi$ equally spaced
azimuths, the real and imaginary parts of its unnormalized Fourier
coefficient each have variance $N_\phi\sigma_0^2/2$.
Denoting the $m$-th Fourier component of the smoothing kernel $g_j(r)$ by
$G_m(r)$, convolution in the azimuthal direction becomes multiplication by
$G_m(r)$ in Fourier space, so that this variance is multiplied by
$|G_m(r)|^2$.
We therefore define $\sigma_C(r)$ as the standard deviation of either the
real or imaginary part of the noise contribution to the Fourier coefficient
$C_m$ defined in Equation~(\ref{eq:complex_fourier_coeff}), giving
\begin{equation}
\sigma_C(r)
=
|G_m(r)|
\sigma_0(r)
\sqrt{\frac{N_\phi}{2}}.
\label{eq:fourier_noise}
\end{equation}

For the phase defined in Equation~(\ref{eq:fourier_phase}), we 
assume negligible covariance between the noise contributions to the real and imaginary parts of $C_m$.
The First-order error propagation then gives
\begin{equation}
\sigma_{\phi_m}(r)
\simeq
\frac{\sigma_C(r)}
{m|C_m(r)|}.
\label{eq:phase_uncertainty}
\end{equation}
Although $N_\phi$ appears explicitly in Equation~(\ref{eq:fourier_noise}),
the phase uncertainty is independent of the azimuthal sampling once the
Gaussian kernel is sufficiently well sampled.
Since $\sum_j g_j^2 \propto N_\phi^{-1}$, we have
$\sigma_0 \propto N_\phi^{1/2}$ and hence
$\sigma_C \propto N_\phi$.
Because the unnormalized Fourier coefficient also scales as
$|C_m| \propto N_\phi$, the $N_\phi$ dependence cancels in
Equation~(\ref{eq:phase_uncertainty}).

At each radius, the pattern speed is obtained from a weighted linear fit to
the phase as a function of time.
We adopt the inverse phase variance,
$w_i=1/\sigma_{\phi_2,i}^2$, as the weight for the $i$-th epoch.
The formal uncertainty in the fitted pattern speed is
\begin{equation}
\sigma_\omega^2(r)
=
\left[
\sum_i
w_i(r)
\left(t_i-\bar{t}_w\right)^2
\right]^{-1},
\qquad
\bar{t}_w
=
\frac{\sum_i w_i t_i}{\sum_i w_i},
\label{eq:omega_uncertainty}
\end{equation}
where $t_i$ is the observing time of the $i$-th epoch and $\bar{t}_w$ is
the weighted mean observing time.

If the phase measurements show excess scatter around the best-fitting
linear relation, such that the reduced chi-square satisfies
$\chi_\nu^2>1$, we inflate the formal uncertainty by a factor of
$\sqrt{\chi_\nu^2}$ and adopt the inflated value as the uncertainty in the
pattern speed.
We note that this uncertainty estimate is based on a simplified treatment of spatially correlated noise. In particular, at radii comparable to or smaller than the beam size, the one-dimensional Gaussian approximation along each ring may become inaccurate. The resulting uncertainties should therefore be regarded as approximate.
\begin{figure}[h]
    \centering
    \includegraphics[width=0.7\linewidth]{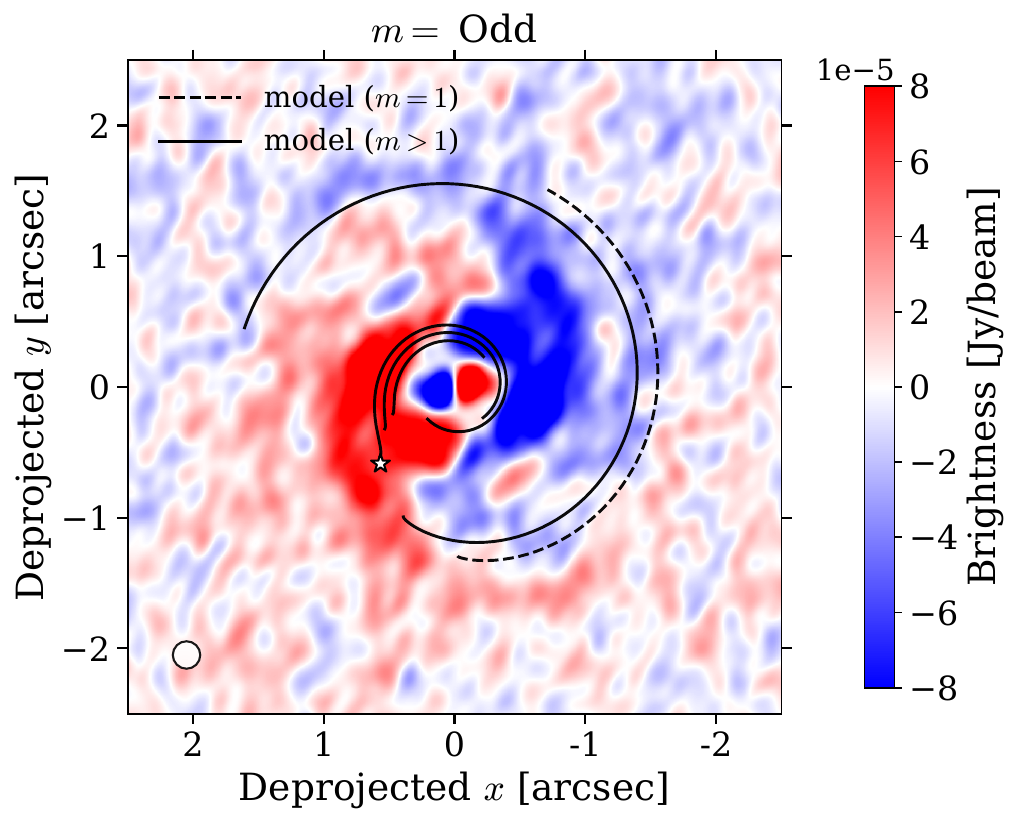}
    \caption{2024 odd-symmetric residual images in \ref{fig:zoom-out_odd} with the planet-driven spiral models \citep{Bae_Zhu2018} overlaid in black. In the panel, the $m = 1$ spiral is shown with a dashed line, while the $m > 1$ spirals are shown with solid lines. In this model, the $m = 1$ spiral appears only exterior to the planet's orbit. The possible planet is located at the position adopted in \citet{Pinte2020}. 
}
    \label{fig:planet_spiral}
\end{figure}

\section{Comparison of the Possible One-Armed Spiral with Linear Planetary Wake Models}
\label{app:Comparison of the Possible One-Armed Spiral}
\citet{Aizawa2026a} suggested that the possible one-armed spiral in Elias~27 could be planet-driven and may have a different origin from the well-known $m=2$ spiral. In contrast to Elias~27, the possible one-armed spiral identified in the IM~Lup disk extends over a much wider radial range than the known $m=2$ spiral.
It also appears to have a larger pitch angle than the nearby northwestern $m=2$ arm, suggesting that the two structures may have different origins.
We therefore compare the observed one-armed spiral with the planet-driven spiral model based on linear density-wave theory \citep{Bae_Zhu2018}, following the approach of \citet{Aizawa2026a}.
The phase of a density wave excited by a planet is given by
\begin{equation}
\phi_{m,n}(R)
=-\phi_{p}
-\mathrm{sgn}(R-R_{p})\frac{\pi}{4m}
+\frac{2\pi n}{m}
-\int_{R^{\pm}_{m}}^{R}
\frac{\Omega(R')}{c_{s}(R')}
\left|
\left\{
1-\left(\frac{R'}{R_{p}}\right)^{3/2}
\right\}^{2}
-\frac{1}{m^2}
\right|^{1/2}
\, dR' .
\end{equation}

Here, $\phi_{m,n}(R)$ is the phase of the planetary wave at radius $R$, $m$ is the azimuthal mode number, and $n$ labels the different wave branches associated with a given $m$. The quantities $R_p$ and $\phi_p$ specify the radial and azimuthal coordinates of the planet, respectively. The inner and outer Lindblad resonance locations are $R^{\pm}_{m}=(1\pm 1/m)^{2/3}R_{p}$, from which the corresponding density waves are launched. The function $\Omega(R)$ and $c_s(R)$ are the local disk angular velocity and sound speed, respectively.

Following \citet{Aizawa2026a} and \citet{Speedie2022}, we assume that the disk rotates at the Keplerian angular velocity, $\Omega(R) =\left({GM_\star}/{R^3}\right)^{1/2}$, and adopt a simple passively irradiated flaring disk model for the thermal structure. The midplane temperature is given by $T_{\rm d}(R)=({\varphi L_\star}/{8\pi R^2\sigma_{\rm SB}})^{1/4}$, where $L_\star$ is the stellar luminosity, $\varphi$ is the disk flaring parameter and $\sigma_{\rm SB}$ is the Stefan--Boltzmann constant.
The local sound speed is then calculated as $c_s(R) = (k_{\rm B}T_{\rm d}(R) / \mu m_{\rm p})^{1/2}$, where $k_{\rm B}$ is the Boltzmann constant, $m_{\rm p}$ is the proton mass, and $\mu$ is the mean molecular weight. We adopt $L_\star=2.6\,L_\odot$, $\varphi=0.02$ and $\mu=2.37$, following
\citet{Speedie2022}, and use $n=0$ to trace the primary planetary wake.
For the planet location, we adopt the midplane position inferred from the CO velocity kink, $R_P =127$~au and $\phi_P=136^\circ$, as compiled by \citet{Speedie2022}.

As shown in Figure~\ref{fig:planet_spiral}, the predicted planetary wakes do not reproduce the morphology of the possible one-armed spiral well.
In particular, the observed spiral appears to have a larger pitch angle than the predicted wakes.
However, the inferred planet location remains uncertain, and a planet at a larger orbital radius could potentially improve the agreement. 

\section{Radius-by-radius Method}
\label{app:radiusbyradius}

We reanalyze the Radius-by-radius method of \citet{Yoshida2025} using the disk geometry inferred in this study. First, we construct deprojected continuum images without subtracting the axisymmetric disk model. These images are produced using the same imaging procedure described in Subsection~\ref{subsec:residual_imaging}, except that CLEAN is performed with \texttt{niter = 50000}.  Following \citet{Yoshida2025}, we calculate the azimuthally averaged intensity profile for each epoch and normalize the images to the 2017 reference image using the ratios of these profiles.

We denote the resulting azimuthal intensity distribution by $I_i(r,\theta)$, where $i$ represents the observing epoch, $r$ is the disk radius, and $\theta$ is the azimuthal angle. Following \citet{Yoshida2025}, we determine the pattern speed at each radius by minimizing the differences between the azimuthal intensity distributions of different epochs. Specifically, we calculate
\begin{equation}
\chi^2(r,\omega)
=
\sum_{i<j}
\sum_{\theta}
\left(\frac{
I_i(r,\theta)
-
I_j(r,\theta-\omega\Delta T_{ij})
}{
\sigma_{ij}
}\right)^2,
\label{eq:radiusbyradius_chi2}
\end{equation}
where $\omega$ is the trial pattern speed, $\Delta T_{ij}$ is the time separation between epochs $i$ and $j$, and $\sigma_{ij}=\sqrt{\sigma_i^2+\sigma_j^2}$, with $\sigma_i$ representing the standard deviation for epoch $i$. We use logarithmically spaced radial bins and adopt the value of $\omega$ that minimizes $\chi^2$ at each radius. The nominal $1\sigma$ uncertainties are determined from the range satisfying $|\chi^2-{\min}(\chi^2)|=1$.
To reduce the effects of differences in the adopted disk centers, we remove the $m=1$ azimuthal Fourier component before measuring the pattern speed. With this treatment, we confirmed that the Keplerian pattern speed profile reported by \citet{Yoshida2025} can be reproduced using the disk centers inferred with \texttt{protomidpy} together with the position angle and inclination from \citet{Huang2018a}.

\section{Linear regression to the Temporal Evolution of the Fourier Phase}
\label{app:linear_regression}
\begin{figure}[h]
    \centering
    \includegraphics[width=1.0\linewidth]{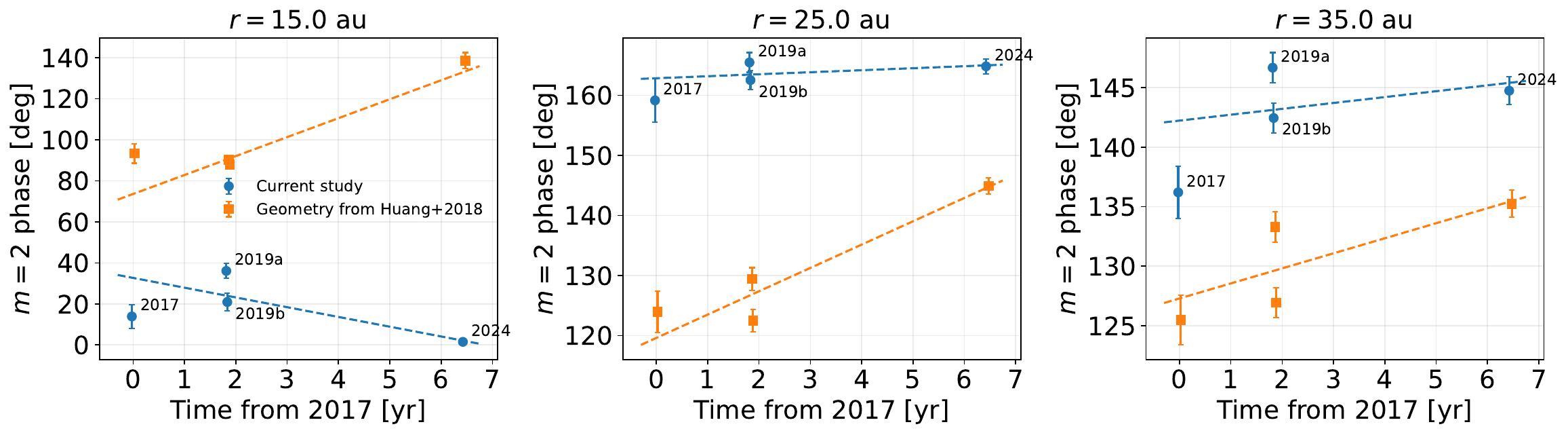}
    \caption{Examples of linear regressions to the temporal evolution of the $m=2$ Fourier phase. The blue circles and orange squares show the phases obtained using the \texttt{protomidpy} and \citet{Huang2018a} geometries, respectively. The dashed lines indicate the corresponding weighted regression line. Positive pattern speeds are defined in the direction of increasing Fourier phase. 
    }
    \label{fig:regression}
\end{figure}

Figure~\ref{fig:regression} presents examples of linear regression to the temporal evolution of the Fourier phase in the inner disk, where our results differ from those of \citet{Yoshida2025}, as shown in the upper panels of Figure~\ref{fig:pattern_speed}. Compared with the fits obtained using the \citet{Huang2018a} geometry, those based on the \texttt{protomidpy} geometry have smaller slopes, corresponding to lower inferred pattern speeds. In the innermost region, including at 15~au, the inferred pattern speeds become negative. The bar-like structure is not clearly recovered in the 2024 Band~7 residual image, suggesting that the Fourier phase may not reliably trace the physical spiral structure in this region. This may contribute to the negative pattern speed inferred there.

\section{Effect of Disk Geometry on the Apparent Spiral Rotation}
\label{app:Reanalysis}

\begin{figure}[htb]
    \centering
    \includegraphics[width=1.0\linewidth]{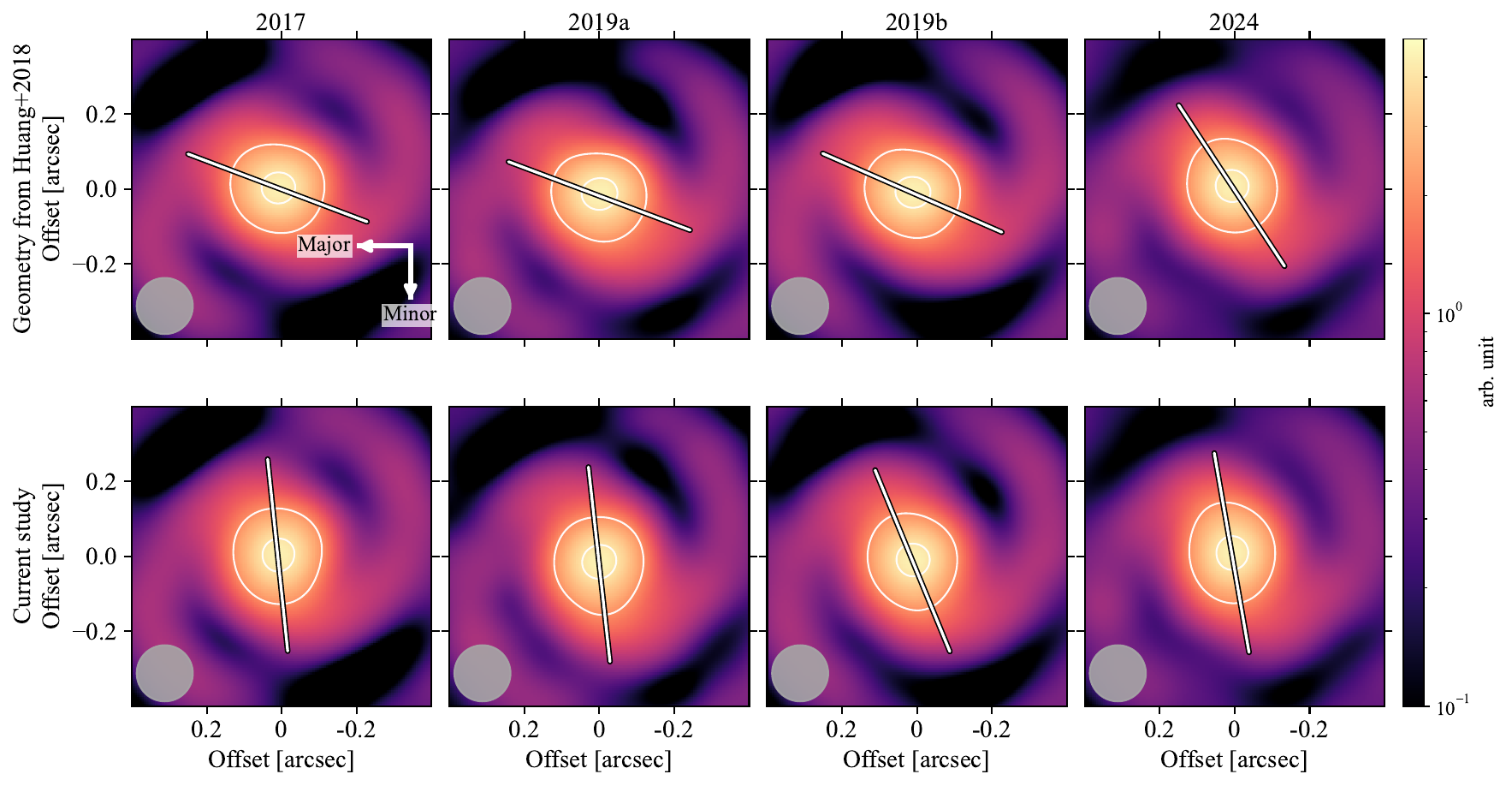}
    \caption{Comparison of the inner disk structures obtained using the geometry from \citet{Huang2018a} (top) and that inferred in this study (bottom), shown for the four epochs. Following \citet{Yoshida2025}, the images are first normalized between epochs, and the non-axisymmetric structure is enhanced by subtracting the azimuthally averaged component from the images after multiplying the original intensity by a factor of 1.2. The white contours indicate levels of 2 and 4 in arbitrary units. The white line overlaid on the contour at a level of 2 indicates the major axis of an ellipse fitted to that contour and is shown as a guide to the orientation of the inner structure.
}
    \label{fig:YoshidaFig3}
\end{figure}

Here, we reanalyze the two approaches in \citet{Yoshida2025} using the disk geometry inferred in this study. 
Figure~\ref{fig:YoshidaFig3} reproduces Figure~3 of \citet{Yoshida2025}, using the two different disk geometries considered in this study. With the \citet{Huang2018a} geometry (top), the inner contrast distribution appears noticeably elliptical in 2017 and 2019 epochs and gives the impression of rotating clockwise toward the 2024 morphology. In contrast, with the geometry inferred in this study (bottom), the inner structures in the Band 6 epochs become more nearly circular and the apparent rotational shift is reduced.
This behavior can be understood as a consequence of geometry mismatch. As shown in Figure~\ref{fig:residual-difference}, an incorrect inclination leaves a characteristic spurious residual pattern along the disk major axis.
In addition, deprojection stretches the image along the project minor axis by a factor of $1/\cos i$.
Because the inclination adopted by \citet{Huang2018a} is smaller than that inferred in this study, the minor axis direction is insufficiently stretched, making the inner structure appear artificially elliptical. The combination of these effects can therefore produce an apparent rotation between epochs. 
\end{CJK*}
\end{document}